\documentclass[journal,comsoc]{IEEEtran}
\IEEEoverridecommandlockouts

\usepackage{graphicx}
\usepackage{float} %tell latex I REALLY want the figure in that specific place, no matter how much whitespace will be left
\usepackage{textcomp} %supports the Text Companion fonts, which provide many text symbols
\usepackage[nospace,noadjust]{cite} %Multiple citations in single cite command
\usepackage{amsmath,amssymb,amsfonts} %IEEE
\usepackage{amsthm} %facilitates the kind of theorem setup
\usepackage{mathrsfs} %support some fonts in mathematics
\usepackage[mathscr]{euscript} %provide spacial capital letters
\usepackage{epstopdf} 
\usepackage[dvipsnames]{xcolor}
\usepackage{pgfplots}
\pgfplotsset{compat=newest} %for double ylabels on the right and the left
\usepackage{tikz}
\usetikzlibrary{plotmarks}
\usepackage[normalem]{ulem} %underlining
\usepackage{gensymb} %Provides generic commands \degree, \celsius, \perthousand, \micro
\usepackage{balance}
\usetikzlibrary{intersections,backgrounds, patterns}
\usepackage{algorithmic}
\usepackage{array}
\usepackage[colorinlistoftodos]{todonotes} %for toddnotes
\usepackage[nolist]{acronym} %for abbreviations
\usepgflibrary{arrows}% for more options on arrows
\usepackage{multirow}

\usepackage[cmintegrals]{newtxmath} %make the symbols of integral, summation and infinity more nice
\usepackage[font=footnotesize]{subcaption}
\usepackage[font=footnotesize]{caption}
\usepackage{algorithm}
\usepackage{algorithmic}

\DeclareMathOperator*{\argmin}{argmin}

\pgfplotsset{every axis/.append style={
		scaled x ticks = false, 
		label style={font=\scriptsize},
		tick label style={font=\scriptsize}
	}
}

\begin{document}

\title{Localization in Ultra Narrow Band IoT Networks: Design Guidelines and Trade-Offs}

\author{Hazem~Sallouha,
        Alessandro Chiumento,
        Sreeraj Rajendran,
        and Sofie~Pollin,
\thanks{Authors are with the Department of Electrical Engineering, KU Leuven, Belgium. e-mail: \{hazem.sallouha@esat.kuleuven.be\}. Alessandro Chiumento is also with CONNECT Centre, Dublin, Ireland.} % <-this % stops a space
}

\begin{acronym}
	
\acro{TD}{Telecom Design}
\acro{$GSN$}{GPS-enabled sensor}
\acro{$SN$}{non-GPS sensor}
\acro{SVM}{support vector machines}
\acro{RndF}{random forest}
\acro{UNB}{ultra narrow band}
\acro{BSs}{base stations}
\acro{RSSI}{received signal strength indicator}
\acro{ML}{machine learning}

\end{acronym}
\maketitle

\begin{abstract}
Localization in long-range Internet of Things networks is a challenging task, mainly due to the long distances and low bandwidth used. Moreover, the cost, power, and size limitations restrict the integration of a GPS receiver in each device. In this work, we introduce a novel received signal strength indicator (RSSI) based localization solution for ultra narrow band (UNB) long-range IoT networks such as Sigfox. The essence of our approach is to leverage the existence of a few GPS-enabled sensors ($GSN$s) in the network to split the wide coverage into classes, enabling RSSI based fingerprinting of other sensors ($SN$s). By using machine learning algorithms at the network backed-end, the proposed approach does not impose extra power, payload, or hardware requirements. To comprehensively validate the performance of the proposed method, a measurement-based dataset that has been collected in the city of Antwerp is used. We show that a location classification accuracy of 80\% is achieved by virtually splitting a city with a radius of 2.5\,km into seven classes. Moreover, separating classes, by increasing the spacing between them, brings the classification accuracy up-to 92\% based on our measurements. Furthermore, when the density of $GSN$ nodes is high enough to enable device-to-device communication, using multilateration, we improve the probability of localizing $SN$s with an error lower than 20\,m by 40\% in our measurement scenario.
\end{abstract}

% Note that keywords are not normally used for peerreview papers.

\IEEEpeerreviewmaketitle

%===========================================================================
\section{Introduction}
%===========================================================================
The Internet of Things (IoT) is becoming ubiquitous, enabling everyday objects to be equipped with computation and communication capabilities \cite{sallouha,antwerpDset,iot,lot18,sallouha2}. To realize IoT, currently deployed networks are using long-range, low power, and low throughput communications such as Sigfox \cite{sigfox}, Weightless-P \cite{wtls}, LoRa \cite{lora}, and narrowband (NB)-IoT \cite{nbIoT}. Among these technologies, \ac{UNB} based solutions, such as Sigfox, offer a compelling mix of simplicity and wide coverage to provide connectivity to millions of devices.

The demand for IoT is set to rise further with the advent of location-based services. In fact, many IoT applications fundamentally depend on the location information to meaningfully interpret any physical measurements collected \cite{iot,lot18}. Unfortunately, the long-range and narrow bandwidth, associated with \ac{UNB} IoT networks, make localization problematic. On one hand, GPS receivers are power hungry and rather expensive to integrate with each IoT node. On the other hand, ranging-based methods lack accuracy because of the long distances and the ultra narrow bandwidth \cite{uwb}. A promising alternative method to tackle these challenges is fingerprinting-based localization. Particularly, by leveraging the advances in \ac{ML} algorithms, fingerprinting methods have expanded to include outdoor scenarios \cite{sallouha,antwerpDset}.

Conventional fingerprinting localization techniques consist of two phases: a training phase, namely an offline phase, and a location estimation phase, namely an online phase. In the training phase, \ac{RSSI} measurements, a.k.a. fingerprints, are collected at known positions and concurrently stored in a dataset. Subsequently, in the online phase, nodes locations are estimated by comparing the real-time \ac{RSSI} measurements with the entries from the dataset \cite{simon}. This comparison process is performed using \ac{ML} algorithms. In general, the main disadvantage of fingerprinting localization is the necessity to keep updating the training datasets. Such a process is both time-consuming and labor-intensive. In this paper, we tackle this drawback by relying on a few \ac{$GSN$} nodes to collect the training data making it, constantly, up to date. This promising potential of fingerprinting methods alongside the insisting demands for an accurate localization technique for \ac{UNB} IoT networks inspired our work in this paper.

\subsection{Related Works}
%===========================================================================
The localization problem has been extensively explored in the literature for various kind of networks in different scenarios \cite{elsawy2018base,win2018theoretical,liu2018mercury,win2011network,win2018network}. However, the new characteristics of IoT networks, such as the long-range, limited number of messages, and the inexpensive IoT nodes, bring extra localization challenges. Therefore, the localization problem in IoT networks has recently attracted considerable research focus \cite{lot18,win2018efficient}. Among all localization techniques, \ac{RSSI}-based methods have been proven as one of the effective solutions \cite{zhang2015nextme,macagnano2014indoor,shit2018location,kwak2018energy,zhang2017path,fonseka2018indoor,song2018csi}. In particular, \ac{RSSI} fingerprinting localization has been widely used in various works \cite{song2018csi,compar,Farjow,kumar,shit2019}, in which both the offline and online phase challenges are addressed.

The main challenge in the offline phase is to minimize the time and effort required to construct a radio map for the area of interest \cite{simon}. To address this issue, the authors in \cite{kumar} and \cite{shit2018probabilistic} adopted probabilistic fingerprinting methods based on RSSI distribution information. Experimental results showed that probabilistic approaches provide good performance. However, the validation of these approaches is limited to indoor environments \cite{shit2019}. For the online phase, the main challenge is mapping the new RSSI measurements to their corresponding locations using the radio map. Here, \ac{ML} algorithms played a significant role in achieving better localization accuracy. A comparison between relevant \ac{ML} algorithms for fingerprinting-based localization is presented in \cite{compar}. Among the presented algorithms, \ac{SVM} and tree-based (e.g., random forest) have shown satisfying accuracy for this range of problems \cite{Farjow,compar}. In \cite{zhou2017device}, the authors proposed SVM based device-free localization using channel state information fingerprinting. Also, in cases where the training dataset and the number of features are relatively small, \ac{SVM} has shown sufficient classification accuracy \cite{Farjow,Tran}. Such characteristics are encountered in long-range networks considered in this work. However, the focus of these aforementioned works is mainly indoor environments. Hence, the proposed methods are incompetent for outdoor long-range IoT networks.

The outdoor localization accuracy of a LoRa deployment using RSSI measurements is investigated in \cite{liy2018towards} where a localization error around 30\,m is achieved. However, the authors considered an optimistic scenario with a very dense gateway deployment. In \cite{xWang} and \cite{sallouha3}, the localization of IoT nodes using mobile anchors is proposed. Although mobile anchors introduce swift and flexible localization solutions, they typically serve limited areas \cite{sallouha3}. A method that uses information about nearby Wi-Fi access points to localize Sigfox nodes is proposed in \cite{janssen2017localization}. This approach achieved location estimation with a median error of 39\,m. However, it requires dedicated Sigfox messages, leading to inefficient use of the limited Sigfox transmissions.

The localization problem in long-range \ac{UNB} IoT networks using RSSI fingerprinting has been addressed in \cite{sallouha,antwerpDset,aernouts}. In \cite{antwerpDset}, the authors introduced datasets collected in large-scale outdoor environments. Moreover, a primary fingerprinting localization using $k$-nearest neighbors ($k$-NN) algorithm on the dataset is examined. A comparison between different \ac{RSSI} based techniques for long-range IoT networks is presented in \cite{aernouts}. The comparison results concluded that the fingerprinting method outperforms other \ac{RSSI} localization methods. However, in both \cite{antwerpDset,aernouts}, the training data has been randomly selected from the whole dataset, which implies having knowledge of \ac{RSSI} fingerprints all over the coverage area. In fact, this requires either a massive number of GPS-enabled nodes to collect tanning data, which is expensive, or alternatively, spend time and effort in building a radio map that is highly vulnerable to become outdated. In \cite{sallouha} a novel measurement based localization approach that leverages the existence of a few GPS nodes (e.g., 1\% of the nodes) is presented. Furthermore, enhancement of the localization accuracy using device-to-device short-range communication is introduced. However, the measurements were only limited to a university campus, and hence, the performance of the proposed approach in large-scale cities is an open problem.

\subsection{Contribution and Paper Structure}
%===========================================================================

In this work, we introduce a novel measurement-based localization approach that leverages the existence of a few \ac{$GSN$} by using their fingerprints to localize other sensors. In particular, we extend the work presented in \cite{sallouha} by applying the localization approach on a larger scale to include an entire city. This allows us to introduce new design parameters and subsequently investigate the trade-offs between them. Our approach starts by classifying the \ac{$SN$} nodes into classes and then followed by a multilateration process within classes for localization with higher precision.

\begin{figure}[t]
	\centering	
	\includegraphics[width=0.45\textwidth]{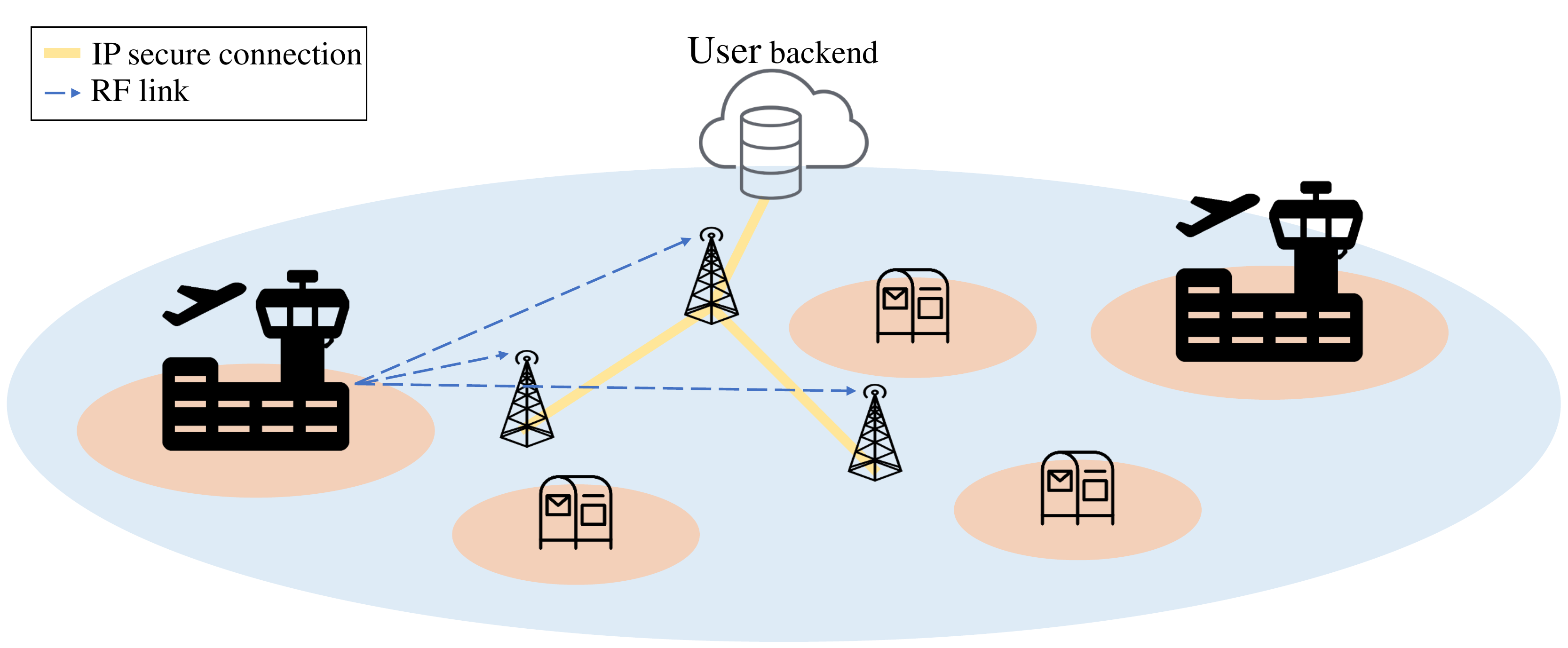} % requires the graphicx package
	\caption{Illustration of a long-range IoT network in which knowing at which post office your package is, provides sufficient location information.}
	\label{iot}
\end{figure}

\subsubsection{Classification based localization} In the first step, we classify \ac{$SN$} nodes into classes anchored by \ac{$GSN$} nodes. For instance, in many applications, such as package localization, knowing in which airport your suitcase is, gives sufficient information. In fact, based on the application requirements, it can be assumed that classes are connected or separated, as shown in Fig. \ref{iot}. To investigate the feasibility of the proposed approach, we introduce various design parameters such as the number of classes, the spacing between adjacent classes, the size of training data, and the number of features used. Each one of these parameters affects the classification accuracy and hence should be chosen carefully based on the application. For instance, the spacing between adjacent classes can be tuned to have localization service all over the coverage area (i.e., connected classes) or limited to certain locations (i.e., separated classes) based on the application requirements.

Moreover, we address the size of the training data, including the number of training examples and the number of features when using \ac{SVM} and \ac{RndF} classifiers. This is particularly important given the limited number of messages in long-range IoT networks. Another essential parameter is the number of \ac{$GSN$}s, represented by the number of classes, assuming at least one \ac{$GSN$} per class. In fact, increasing the density of \ac{$GSN$}s will result in smaller classes radii, lower localization errors, but more similarity in the fingerprints of nearby \ac{$GSN$}s and hence higher classification error. Therefore, in the case of high \ac{$GSN$}s density, we can rely on device-to-device (D2D) communication to perform multilateration.

\subsubsection{Regression-based distance estimation} This step improves the localization accuracy by relying on device-to-device (D2D) communication between the \ac{$GSN$} anchors and \ac{$SN$}s, as IoT nodes, typically, by default, have modems that support short-range D2D communication \cite{TD}. Instead of decreasing the localization errors by increasing the number of classes, i.e., \ac{$GSN$}s, we fine-tune the location within the class by performing multilateration using three \ac{$GSN$}s. We can then localize nodes in a two-step method: first, fingerprinting classification using \ac{$GSN$}s and then multilateration relying on D2D communication. In fact, limiting the use of \ac{RSSI} distance estimation to short distances provides satisfactory performance.

In contrast to the existing literature on fingerprinting localization, our proposed two-step method saves time and effort required to build a full radio map. Such a radio map is highly vulnerable to become outdated, particularly with the long-range and outdoor deployment considered. Instead, we first rely on \ac{$GSN$} anchors, with their constantly updated \ac{RSSI} measurements, to classify \ac{$SN$}s. However, since the locations of \ac{$GSN$}s are sparse, the estimated location of a given \ac{$SN$} node is represented by a zone, namely, a class. Such an estimated zone is, in fact, sufficient for many applications, such as package localization. Nevertheless, to further refine this estimation, we introduce the second step for those applications that require more localization precision. This step improves the precision, conditioned on the availability of D2D communication. In particular, the second step relies on regression-based distance estimations, delivering the exact position of the \ac{$SN$} node.

The rest of the paper is organized as follows. Section \ref{backStrc} details the communication technologies used and clarifies the localization problem in long-range \ac{UNB} IoT networks. In Section \ref{fbClass}, we introduce our fingerprinting base approach and detail the \ac{ML} algorithms considered in this work. Localization based on distance estimation is presented in Section \ref{regSection}. Our experimental results using a real Sigfox deployment are presented in Section \ref{results}. Finally, we conclude our work in Section \ref{conc}.

%===========================================================================
\section{Background and Problem Statement}\label{backStrc}
%===========================================================================
In this section we introduce the structure of \ac{UNB} IoT network. Subsequently, the localization problem in such networks is thoroughly detailed.

\subsection{Background}
%===========================================================================

Sigfox \cite{sigfox}, is an IoT provider which uses \ac{UNB} channels. The physical layer of Sigfox uses BPSK modulation technique providing a fixed bit rate of 100 bps with a payload up to 12 bytes. Moreover, the ISM 868 MHz band is used from which a bandwidth of $40$kHz is split, providing orthogonal channels with a bandwidth of 100Hz each. Consequently, IoT sensor nodes randomly pick a channel for each transmission \cite{brecht}. Following the regulations of the ISM band, Sigfox MAC layer is designed to limit the number of transmissions to 140 messages a day.

\subsubsection{Sigfox - Cellular IoT Network}
%===========================================================================
Sigfox is a star cellular network in which the uplink data flow from nodes to \ac{BSs} are assumed to be 97\% of the overall traffic. An illustration of a Sigfox network is shown in Fig. \ref{iot}. As one can see in the figure, the network consists of three parts: 1) the things, which could be any sensor or package equipped with an RF transmitter, 2) the base stations, which collect the data transmitted by the sensors and 3) the network's cloud to which all base stations are connected. Moreover, the user back-end could be added optionally, in different forms, to allow the users to access their data either via email, API, or store it to their database. In the scope of this paper, we highlight the three main elements used in our experiments.

\begin{itemize}
\item \textit{Base stations (BS)s}: A single Sigfox base station can provide a coverage range up to 40\,km. Hence few \ac{BSs} are able to cover an entire city, both indoor and outdoor \cite{sigfox}. Sigfox \ac{BSs} are connected to a centralized user back-end, enabling connectivity for millions of nodes.

\item \textit{Sensor Nodes (SN)}: The sensor nodes used are equipped with \ac{TD} modems (TD1208). These modems support communication over Sigfox and use a transmit power up to 25\,mW. 
\item \textit{GPS-enabled nodes (GSN)}: They are nodes with the same functionality as \ac{$SN$}. However, their modems are equipped with a GPS-receiver (TD1204) and, therefore, their coordinates are included in the messages they send.
\end{itemize}
Another possible implementation is to use hybrid networks in which \ac{$SN$} nodes forward data to gateway nodes which, in turn, send it to the \ac{BSs} \cite{TD}. Such a short relay connection can be made using \ac{TD}-LAN. In our case, we assume that \ac{$GSN$}s receive messages from \ac{$SN$}s. This LAN network, with its short range communication, can be used to increase the localization accuracy of \ac{$SN$} nodes within classes.
\subsubsection{Device-to-Device LAN enabled}
%===========================================================================
Nodes used in our experiments have \ac{TD} Sigfox modems. These modems support communication over both Sigfox and \ac{TD} LAN. Therefore, device-to-device communication can be enabled to build a \ac{TD}-LAN network \cite{TD}. The \ac{TD}-LAN is an energy-efficient network uses time division duplex (TDD) with narrow bandwidth channels of 25 kHz. Gaussian minimum shift keying (GMSK) is used as a modulation technique, and the frequency band is similar to Sigfox, i.e., ISM 869 MHz. In \ac{TD}-LAN, the payload can go up to 17 bytes with a transmission rate of 9600 bps. A star topology can be formed by fixing a particular node in receive mode while its neighbors transmit upon request.

\subsection{Problem Statement}
%===========================================================================
Each Sigfox message is received by all reachable \ac{BSs}. However, they are typically far, in a region where the \ac{RSSI}-distance curve resolution is not sufficient. Since long-range IoT networks are cellular based, one might intuitively suggest using the multiple \ac{BSs} to perform multilateration ranging-based techniques for localization. In fact, unlike the typical mobile cellular networks, in long-range IoT networks, each base station covers a relatively large distance, i.e., up to 40\,km \cite{brecht,sigfox}, hence leading to a rather large uncertainty zone. Particularly, in multilateration process, the accuracy depends on the distance from nodes to \ac{BSs}. This distance is estimated using time of arrival (ToA) or \ac{RSSI}. Firstly, if ToA is used, the Cram\'er-Rao lower bound (CRLB) of the estimated distance $\hat{d}$ is given by \cite{uwb}
\begin{eqnarray}
\sqrt{\text{Var}(\hat{d})}\geqslant \frac{c}{2\sqrt{2} \pi \sqrt{SNR} \beta}\,,
\label{TOA_LB}
\end{eqnarray}
where $Var\lbrace \cdot \rbrace$ is the variance, $c$ is the speed of light, $SNR$ is the signal-to-noise ratio, and $\beta$ is the effective signal bandwidth. Based on (\ref{TOA_LB}), the accuracy of ToA localization techniques is directly proportional to the signal bandwidth. Long-range communication mainly works with \ac{UNB} transmission \cite{sigfox}. Therefore, the \ac{UNB} channels of 100Hz lead to a significantly low time resolution in which the transmission time goes up to 1-2 seconds per message. Such a low time resolution severely degrades the estimation accuracy.

\begin{figure}[t]
	\centering
	% This file was created by matlab2tikz.
%
%The latest updates can be retrieved from
%  http://www.mathworks.com/matlabcentral/fileexchange/22022-matlab2tikz-matlab2tikz
%where you can also make suggestions and rate matlab2tikz.
%
\definecolor{mycolor1}{rgb}{0.00000,0.44700,0.74100}%
\definecolor{mycolor2}{rgb}{0.85000,0.32500,0.09800}%
\definecolor{mycolor3}{rgb}{0.92900,0.69400,0.12500}%

\begin{tikzpicture}

\begin{axis}[%
legend style={font=\fontsize{6}{5}\selectfont},
width=6cm,
height=4.3cm,
at={(1.154in,0.752in)},
scale only axis,
xmin=-105,
xmax=-70,
xmajorgrids,
xlabel={RSSI (dBm)},
ymin=0,
ymax=40,
ymajorgrids,
ylabel={count},
axis background/.style={fill=white},
legend style={at={(0.03,0.97)},anchor=north west,legend cell align=left,align=left,draw=white!15!black}
]

\addplot[fill=mycolor2,fill opacity=0.7,draw=black,ybar interval,area legend] plot table[row sep=crcr] {%
x	y\\
-72.5	31\\
-71.5	29\\
-70.5	29\\
};
\addlegendentry{$d$ = 10m};

\addplot[fill=mycolor2,fill opacity=0.7,draw=black,ybar interval,area legend,pattern color = mycolor2, pattern = north east lines] plot table[row sep=crcr] {%
x	y\\
-77.5	5\\
-76.5	36\\
-75.5	19\\
-74.5	19\\
};
\addlegendentry{$d$ = 30m};

\addplot[fill=mycolor3,fill opacity=0.7,draw=black,ybar interval,area legend] plot table[row sep=crcr] {%
x	y\\
-87.5	14\\
-86.5	14\\
-85.5	21\\
-84.5	11\\
-83.5	11\\
};
\addlegendentry{$d$ = 60m};

\addplot[fill=mycolor1,fill opacity=0.7,draw=black,ybar interval,area legend,pattern color = mycolor1, pattern = north west lines] plot table[row sep=crcr] {%
x	y\\
-93.5	2\\
-92.5	16\\
-91.5	14\\
-90.5	22\\
-89.5	6\\
-88.5	6\\
};
\addlegendentry{$d$ = 80m};

\addplot[fill=mycolor1,fill opacity=0.7,draw=black,ybar interval,area legend] plot table[row sep=crcr] {%
x	y\\
-102.5	1\\
-101.5	11\\
-100.5	14\\
-99.5	24\\
-98.5	7\\
-97.5	3\\
-96.5	3\\
};
\addlegendentry{$d$ = 130m};

\end{axis}
\end{tikzpicture}%
	\caption{A Histogram showing the variance of RSSI measurements at various distances.}
	\label{point}
\end{figure}
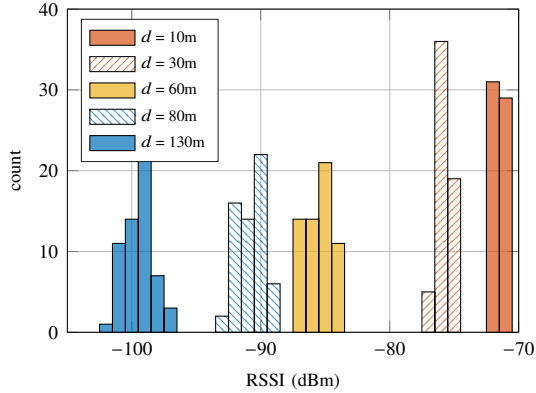

For the case of \ac{RSSI}, the CRLB of an estimated distance $\hat{d}$ is given by the following inequality \cite{uwb}
\begin{eqnarray}
\sqrt{\text{Var}(\hat{d})}\geqslant \frac{\ln 10}{10}\frac{\sigma_{sh}}{n_p}d\,,
\label{CLrss}
\end{eqnarray}
where $d$ is the distance between the node and the base station, $n_p$ is the path loss exponent, and $\sigma_{sh}$ represents the standard deviation of the log-normal shadowing effect. By detailing the variables in (\ref{CLrss}), we first have $\sigma_{sh}$ and $n_p$ that are channel parameters and hence, are out of control for a given environment. In (\ref{CLrss}), moreover, one observes that the estimator accuracy is inversely proportional to the distance $d$, i.e., the larger the distance, the lower the accuracy. This means that poor accuracy is achieved with long-range communication.

\begin{figure}[t]
	\centering	\includegraphics[width=0.45\textwidth]{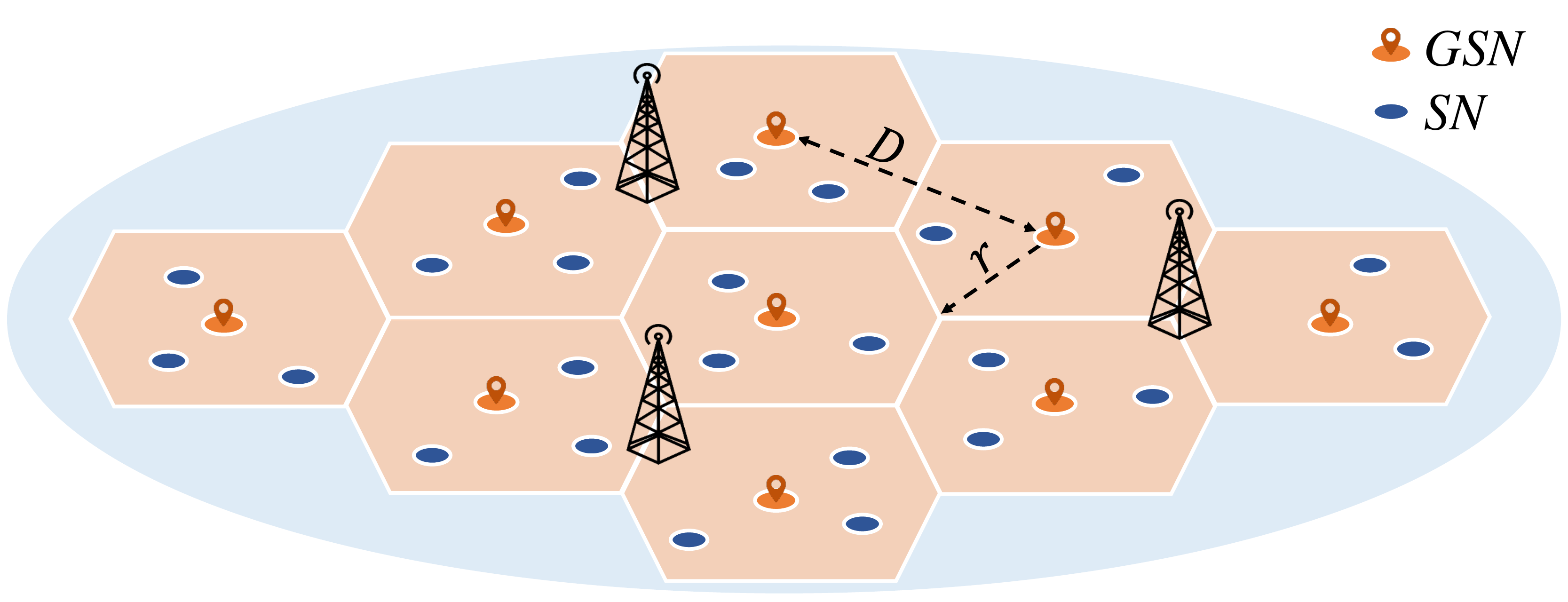} % requires the graphicx package
	\caption{Classifying \ac{$SN$}s to one of the virtual \ac{$GSN$}'s classes.}
	\label{model}
\end{figure}

Interestingly, \ac{RSSI} measurements present two distinctly different behaviors when the receiver is close to the transmitter, rather than further away \cite{sallouha}. For short distances, the \ac{RSSI}-distance curve decreases rapidly with the distance. However, as the distance increases ($>$100\,m), the \ac{RSSI}-distance slope decreases asymptotically. Concurrently, the variations around the mean increases, as shown in Fig. \ref{point}. Therefore, \ac{RSSI} measurements can be used for distance estimation and fingerprinting, depending on the distance. At short distances, ranging using a regression process gives an accurate estimation of the distance between the node and the base station. On the other hand, as the distance increases and hence, the variance of the \ac{RSSI} measurements, the distance estimation error rises dramatically. Therefore, it is preferable to leverage such variations to distinguish between different locations and subsequently, classify nodes into classes using their RSSI measurements.

%===========================================================================
\section{Fingerprinting Based Classification}\label{fbClass}
%===========================================================================
In this section, we present our novel long-range localization approach based on region partitioning using \ac{$GSN$}s. In particular, \ac{$GSN$}'s \ac{RSSI} fingerprints are used as landmarks (i.e., training data) for our classification framework, allowing us to divide the \ac{$SN$}s into classes. In the following, we introduce our system design. Subsequently, we highlight two algorithms which are widely used for such classification problems: \ac{RndF} and \ac{SVM} classification algorithms. Finally, we present our algorithm summary and detail the error metric of our classification problem.

\subsection{System Design}
%===========================================================================

Consider an outdoor \ac{UNB} IoT network in which sensor nodes are randomly distributed in a wide region. Assume two kinds of sensor nodes: \ac{$GSN$} and \ac{$SN$}. We aim at using the nodes $GSN_1$, $GSN_2$, $\ldots$, $GSN_L$ with their know locations $c_1$, $c_2$, $\ldots$, $c_L$, respectively, to split the wide region into $L$ classes, as shown in Fig. \ref{model}. The classes are assumed to have equal radii denoted as $r$ with \ac{$GSN$}s located at the centers. Moreover, we assume equal spacing $D$ between the centers of adjacent classes, i.e., adjacent \ac{$GSN$}s. In order to split the region into classes based on \ac{$GSN$}s, we have $D \geq \sqrt{3}r$. Accordingly, the spacing between adjacent classes, $x$, is defined as:
\begin{eqnarray}
x = D - \sqrt{3}r,\,\,\,  & x \in [0,D)\,.
\end{eqnarray}
If \ac{$SN$}s are uniformly distributed all over the region then, we choose classes with $D = \sqrt{3}r$, i.e., $x = 0$. However, if \ac{$SN$}s are located in given parts of the region, then we choose \textit{separated classes} with spacing $x >0$, i.e., $D > \sqrt{3}r$.

\begin{table}[t!]
	\small
	\vspace{-0.5em}
	\caption{Representation of RSSIs collected from one \ac{$GSN$} node}
	\label{rssT}
	\centering
	\begin{tabular}{ccccc}
		\firsthline \hline
		\multicolumn{5}{c}{\hspace{2cm}RSSI (dBm)} \\
		\cline{2-5}
		{Time index}  & BS$_{1}$  & BS$_{2}$ &\ldots  & BS$_{M}$\\ 
		\hline
		t = 0 & RSSI$_1^{(0)}$ & RSSI$_2^{(0)}$ & \ldots & RSSI$_M^{(0)}$   \\ [1ex]
		t = 1 & RSSI$_1^{(1)}$ & RSSI$_2^{(1)}$ & \ldots & RSSI$_M^{(1)}$   \\ [1ex]
		\vdots          & \vdots		   & \vdots        &$\ddots$	& \vdots 	  \\ [1ex]
		t = $T-1$ & RSSI$_1^{(T-1)}$ & RSSI$_2^{(T-1)}$ &\ldots & RSSI$_M^{(T-1)}$   \\ [1ex]
		\lasthline
	\end{tabular}
\end{table}

In long-range UNB IoT networks, each message sent by \ac{$GSN$} or \ac{$SN$} is received by $M$ base stations at which \ac{RSSI} is measured. This gives us a vector \textbf{R}$^{(t)}$ = [RSSI$_1^{(t)}$, RSSI$_2^{(t)}$, ..., RSSI$_M^{(t)}$] $\in \mathbb{R}^{M}$, with the element \ac{RSSI}$_m^{(t)}$ representing the \ac{RSSI} sample recorded at time index $t$ at base station $m$, where $t = 0,1,...,T-1$ and $m = 1,2,...,M$. Having $T$ messages received, it is then possible to construct a matrix \textbf{C} = [\textbf{R}$^{(0)}$, \textbf{R}$^{(1)}$, ..., \textbf{R}$^{(T-1)}$]$^{T}$ $\in \mathbb{R}^{T\times M}$, with [.]$^{T}$ being the transpose operator.

Now, using $L$ \ac{$GSN$}s, we assume that the region is divided into $L$ classes based on $GSN_1$, $GSN_2$, $\ldots$, $GSN_L$ with their matrices \textbf{C}$_1$, \textbf{C}$_2$, ..., \textbf{C}$_L$, respectively. Table \ref{rssT} presents the \ac{RSSI} values of matrix \textbf{C}$_l$ for a given \ac{$GSN$}, namely $GSN_l$. Considering all $L$ classes, the measurements can be stored in a three-dimensional matrix with a size of $T \times M \times L$. In classification problems, this represents a $T \times L$ set of training examples with $M$ different features. Now, given the locations and the \ac{RSSI} fingerprints collected from \ac{$GSN$}s, we proceed to classify all \ac{$SN$}s to one of these classes using \ac{ML} algorithms. In the next section, we present the relevant algorithms considered in this work.

\subsection{Supported Vector Machines}
%===========================================================================

\ac{SVM} is a supervised \ac{ML} algorithm commonly used in classification problems. It is distinguished by its support vectors and kernel function \cite{Tran}. The support vectors are chosen in the training phase and are defined as the critical elements of the dataset. This set of critical elements determines a \textit{separating hyperplane} with the objective of maximizing decision margins. The learning of the hyperplane is done by mapping the dataset into a higher dimension. This is where the kernel plays a role. In this paper, we consider the RBF kernel function, a.k.a. Gaussian kernel. RBF kernel is known for its empirical effectiveness and is easy to calibrate \cite{kumar,Farjow}. Now, let vector $\textbf{R}$ denotes a training example from a given class, i.e., \ac{$GSN$} node, and vector $\textbf{R}'$ indicates a test example from a given \ac{$SN$} node, the corresponding RBF kernel is defined as 
\begin{eqnarray}
\mathcal{K}({\textbf{R}},{\textbf{R}'})&=&\exp \Big(-\frac{\lVert {\textbf{R}}-{\textbf{R}'}\lVert^2}{2\sigma^2} \Big)\,,
\end{eqnarray}
where $\lVert.\lVert$ is the euclidean distance. $\sigma^2$ is known as the smoothness parameter of the kernel (i.e., it controls the influence of individual training examples). In other words, low $\sigma^2$ means that only \textit{border examples} are used in determining the separating hyperplane. Whereas with high values of $\sigma^2$ \textit{far from the border examples} are also considered in defining the separating hyperplane. It is worth noting that, in our implementation, $\textbf{R}$ and $\textbf{R}'$ represent vectors of \ac{RSSI} measurements collected from \ac{$GSN$} and \ac{$SN$}, respectively.

To gain insight into how to select the value of $\sigma^2$, consider the \ac{RSSI} histogram shown in Fig. \ref{point}. For short distances (e.g., 10\,m and 30\,m), where \ac{RSSI} varies sharply, a kernel function with low $\sigma^2$ ($\approx0.01$) is needed. On the contrary, at long distances where \ac{RSSI} varies smoothly with the distance, a kernel function with high $\sigma^2$ ($\approx 2$) is preferable. In fact, high and low values of $\sigma^2$ lead to \ac{SVM} with simple and sophisticated separating hyperplane, respectively. Another important parameter that should be considered when choosing the value of $\sigma^2$, is the number of features, $M$. In fact, a favorable initial value of $\sigma^2$ is $M/2$ \cite{scikit}. Subsequently, we tune it, if needed, based on the way the training dataset varies.

\subsection{Random Forest}
%===========================================================================
\ac{RndF} is ensemble classifier based on creating various tree predictors in the training phase and then output the class labels which have the majority vote among the trees. Each tree is grown by randomly drawing samples, with replacement, from the training set. In our localization approach messages collected from \ac{$GSN$} are used to construct trees. These trees are then combined by averaging their probabilistic prediction. \ac{RndF} attains high classification accuracy and can handle outliers and noise in the data \cite{ahmad2018performance} particularly, in cases with a rather high number of features. Moreover, it is known for being less susceptible to over-fitting.

\subsection{Algorithm Summary and Error Metric}
%===========================================================================
The steps for updating the training data and estimating the class of any given \ac{$SN$} are summarized in Algorithm \ref{alg:class}. The algorithm works as follows. When a new message arrives at the network back-end, it first asks for the RSSI measurements from all \ac{BSs}, i.e., \textbf{R}$^{(t)}$. Subsequently, the algorithm checks the message sender. If the sender is one of the \ac{$GSN$}s, it uses \textbf{R}$^{(t)}$ to update the corresponding class matrix, namely, \textbf{C}$_l$ (lines 7-9). Otherwise, the message is assumed to be from an \ac{$SN$}. Hence, the algorithm proceeds to estimate its class denoted as $SN_\text{class}$. The \ac{ML} model is trained periodically, as described in Algorithm \ref{alg:class} (lines 13-15). In particular, the \ac{ML} model's retraining interval is defined as \textit{trainPeriod}. This period depends on the message transmission rate and the environment. A reasonable empirical value for a Sigfox network in an urban area is one hour. Note that, in Algorithm \ref{alg:class}, the number of rows in the training data, $T$, is fixed. This means that whenever a new training message is recorded, it overwrites its oldest counterpart, thus keeping the training dataset updated.

\begin{algorithm} [t]
	\caption{Model training and classification of $SN$s}
	\label{alg:class}
	\begin{algorithmic}[1]
		\STATE \textbf{Input:} $GSN_1$, $GSN_2$ $\ldots$, $GSN_L$\\ 
		ML $\in$ \{SVM, RndF\} and the required number of rows in the training data $T$
		\STATE \textbf{Output:} Estimated class of the $SN$: $SN$$_\text{class}$
		\item[]
		\WHILE{True}
		\STATE newMessage $\leftarrow$ receiveMessage\,(\,)
		\STATE \textbf{R}$^{(t)}$ $\leftarrow$ newMessage.measureRSSI\,(\,)
		\IF {newMessage.sender $\in$ $GSN_1$, $GSN_2$ $\ldots$, $GSN_L$}
		\STATE \textbf{C}$_l$ = newMessage.getC\,(\,)
		\STATE \textbf{C}$_l[\,t_l\,\%\,T\,,\,:\,]$ $\leftarrow$ \textbf{R}$^{(t)}$ \,\,\,\,\,\,{\#}\,$t_l$ is the time index for \textbf{C}$_l$
		\STATE $t_l \leftarrow t_l+1$
		\ELSE
		\STATE $SN$$_\text{class}$ = ML.predict\,(newMessage) \,\,\#\,classification
		\ENDIF
		\item[] \#\,training phase
		\IF{time.seconds\,(\,) \% trainPeriod $ == 0$}
		\STATE ML.train\,(\textbf{C}$_1$,  $\ldots$, \textbf{C}$_l$ $\ldots$, \textbf{C}$_L$)
		\ENDIF
		\ENDWHILE
	\end{algorithmic}
\end{algorithm}

As shown in Algorithm \ref{alg:class}, the \ac{$SN$} in an unknown class centered at $c$ is classified into one of the \ac{$GSN$} known classes. Therefore, one can define the classification error as
\begin{eqnarray}
\boldsymbol{\xi}&=&  {\lVert\hat{c}} - {{c}}\rVert^2\,,
\label{classError}
\end{eqnarray}
where $\hat{c}$ is the center of the estimated class. Consequently, in case a classification error occurs, the minimum error is defined by the distance $D$ between adjacent classes (i.e., the separation between adjacent \ac{$GSN$}s). Consequently, one can write
\begin{eqnarray}
\min \hspace{0.1cm}\boldsymbol{\xi}&=&  \min{\lVert\hat{c}} - {{c}}\rVert^2  \hspace{0.2cm}= \hspace{0.2cm} D\,.
\label{classMinE}
\end{eqnarray}
Now, define the radius $r$ as the distance between \ac{$GSN$} and the furthest \ac{$SN$} in a given class. The classification error decreases when $x$ increases, i.e., when $ D\gg \sqrt{3}r$. The influence of $x$ on the classification accuracy, for a given number of classes, is presented in the results section using measurements.

%===========================================================================
\section{Regression-based Distance Estimation}\label{regSection}
%===========================================================================

Exploiting the benefits of having \ac{$GSN$}s can be extended by using short-range device-to-device (D2D) communication between \ac{$GSN$} and \ac{$SN$} nodes. If three \ac{$GSN$}s are available within the communication range, \ac{RSSI} values can be used for distance estimation \cite{simon,kumar}. Thus, in addition to classifying \ac{$SN$}s into classes, it is possible to further estimate their location within the class. This can be achieved by estimating distances between \ac{$SN$} and \ac{$GSN$}s. Even though this step requires at least 3 \ac{$GSN$}s, it can substantially increase the localization accuracy.

\subsection{Regression}

Instead of building a full radio map, the regression process is a sufficient alternative to estimate distances as a function of \ac{RSSI}. Basically, a regression process interpolates discrete measurements and generates a continuous output function, which subsequently used to map any \ac{RSSI} value to its corresponding distance \cite{Chiumento,Chiumento2015}. Assume \ac{RSSI} samples are collected at $J$ different distances from a given \ac{$GSN$} node, the regression problem can be written as
\begin{eqnarray}
\argmin_{\boldsymbol{a}} \bigg\{  \Big( \sum_{j=1}^{J} {\varPsi} ({{RSSI}_j}) - {d}_j \Big)^2 \bigg\}\,,
\label{regrProb}
\end{eqnarray}
where $d_j$ is the distance of the $j$-th location from the \ac{$GSN$}. Assuming an $n$-th degree polynomial regression function, ${\varPsi} ({{RSSI}_j})$ is expressed by
\begin{eqnarray}
{\varPsi} ({{RSSI}_j})
&=& a_0 + a_1{RSSI}_j^{} + a_2{RSSI}_j^{2} + \nonumber\\
 &\cdots& + a_n{RSSI}_j^{n}\,,
\label{regrPoly}
\end{eqnarray}
where $a_i$ (with $i = 0, 1, 2, ..., n$) is the $i$-th coefficient and $j = 1,2,...,J$. The motivation behind choosing the polynomial regression is its evaluation simplicity and the fact that polynomials dominate the interpolation theory \cite{impRss}. Once ${\varPsi} ({{RSSI}_j})$ is defined, the distance estimated from a new RSSI$_t$ measurement is written as
\begin{eqnarray}
{\hat{d}} &=& \argmin_{d_j} \bigg\{  \Big( {\varPsi}^{-1} (d_j) - {RSSI}_t \Big)^2 \bigg\}\,,
\label{regrDis}
\end{eqnarray}

Here, it is worth noting that ${\varPsi}^{-1} (d_j)$ is a continuous function. Hence, $\hat{d}$ can take any non-negative value.

\begin{algorithm} [t]
	\caption{Localizing $SN$s with D2D communication}
	\label{alg:loc}
	\begin{algorithmic}[1]
		\STATE \textbf{Input:} $GSN_1$, $GSN_2$ $\ldots$, $GSN_L$ with their locations $c_1$, $c_2$, $\ldots$, $c_L$, respectively\\
		The $SN$ to be localized
		\STATE \textbf{Output:} Estimated location of the $SN$: $SN$$_\text{location}$
		\item[]
		\STATE $K \leftarrow \sum_{i=1}^{L} \,\,$ inRange\,(\,$SN$,$GSN_i$)
		\IF{$K \geq 3$}
		\STATE Use (\ref{regrDis}) to compute $\hat{d_1}, \hat{d_2}, \ldots, \hat{d}_K$
		\STATE $SN$$_\text{location}$ = multilateration($\hat{d_1}, \hat{d_2},...,\hat{d}_K,c_1,c_2,...,c_K$)
		\ELSE
		\STATE $SN$$_\text{location}$ = $SN$$_\text{class}$ \hspace{1cm} \#\,from Algorithm 1
		\ENDIF
	\end{algorithmic}
\end{algorithm}

\subsection{Algorithm Summary and Error Metric}
%===========================================================================
Algorithm \ref{alg:loc} presents the details of estimating the location of a given $SN$ when D2D communication is enabled. Once $SN$$_\text{class}$ is estimated using Algorithm \ref{alg:class}, the position within the class, $SN$$_\text{location}$, can be estimated using Algorithm \ref{alg:loc}. Let $K$ denotes the number of $GSN$ nodes which are in \textit{range} of the target $SN$. The function \textit{inRange}($SN$, $GSN_i$) simply returns one if $GSN_i$ is able to receive messages from $SN$ and zero otherwise. If $K \geq 3$, the algorithm uses (\ref{regrDis}) to find $\hat{d_1}, \hat{d_2}, \ldots, \hat{d}_K$ and subsequently performs the multilateration process. However, if $K < 3$, the multilateration process is not applicable. Hence, it returns $SN$$_\text{class}$ estimated in Algorithm \ref{alg:class}.

In practice, the function ${\varPsi} ({{RSSI}_j})$ does not fit the data perfectly. Therefore, an estimation error is expected. The average localization error is given by
\begin{eqnarray}
{\xi}&=& \sqrt{\frac{1}{KT_s}\sum_{t=1}^{T_s} \sum_{k=1}^{K} {{| \hat{d}_{tk}} - {d_{tk}} |}^2}\,,
\label{regrRMSE}
\end{eqnarray}
where $T_s$ is the number of RSSI measurements used to estimate the distance between one $GSN$ and the given $SN$ node. Moreover, $\hat{d}_{tk}$ in (\ref{regrRMSE}), represents the estimated distance from the $k$-th $GSN$ using the $t$-th RSSI measurement.

\begin{figure}[t]
	\centering
	\includegraphics[width=0.33\textwidth]{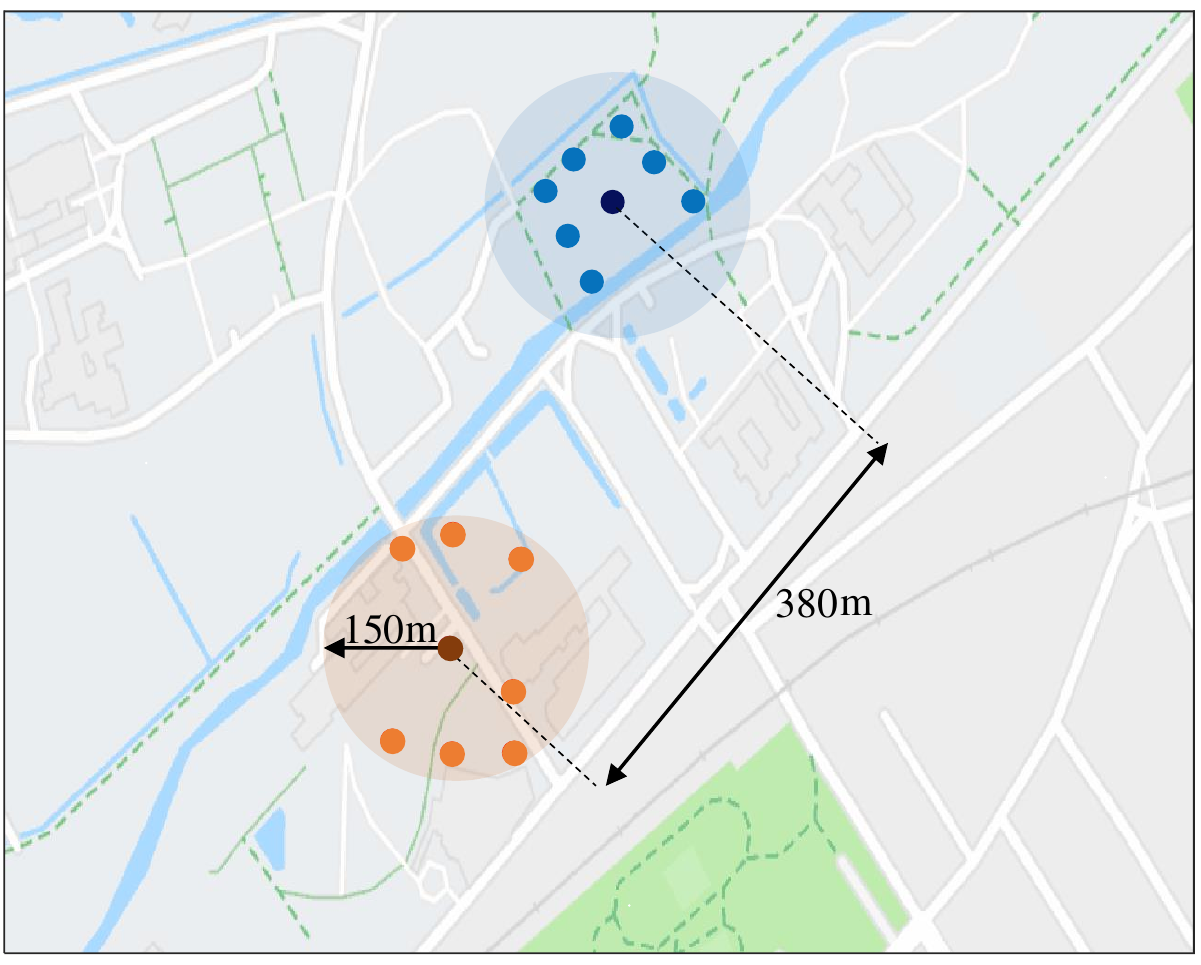}
	\caption{A map showing the nodes' positions in the university campus with $D \approx \frac{3\sqrt{3}}{2}r$. Dark orange and dark blue are \ac{$GSN$}s.}
	\label{map}
\end{figure}

\begin{figure*}[t]
	\begin{subfigure}[b]{0.33\textwidth}
		\includegraphics[width=\textwidth]{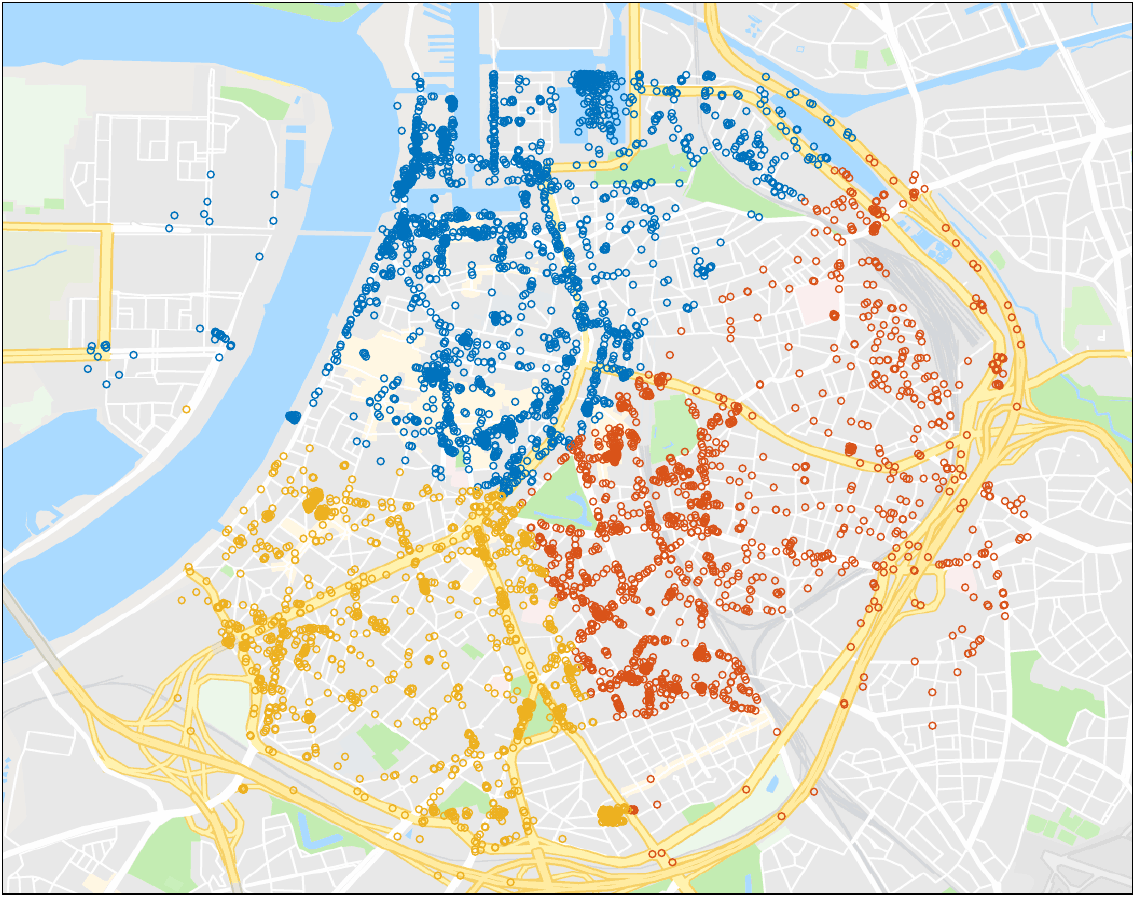}
		\caption{Three classes}
		\label{3cmap}
	\end{subfigure}
	\begin{subfigure}[b]{0.33\textwidth}
		\includegraphics[width=\textwidth]{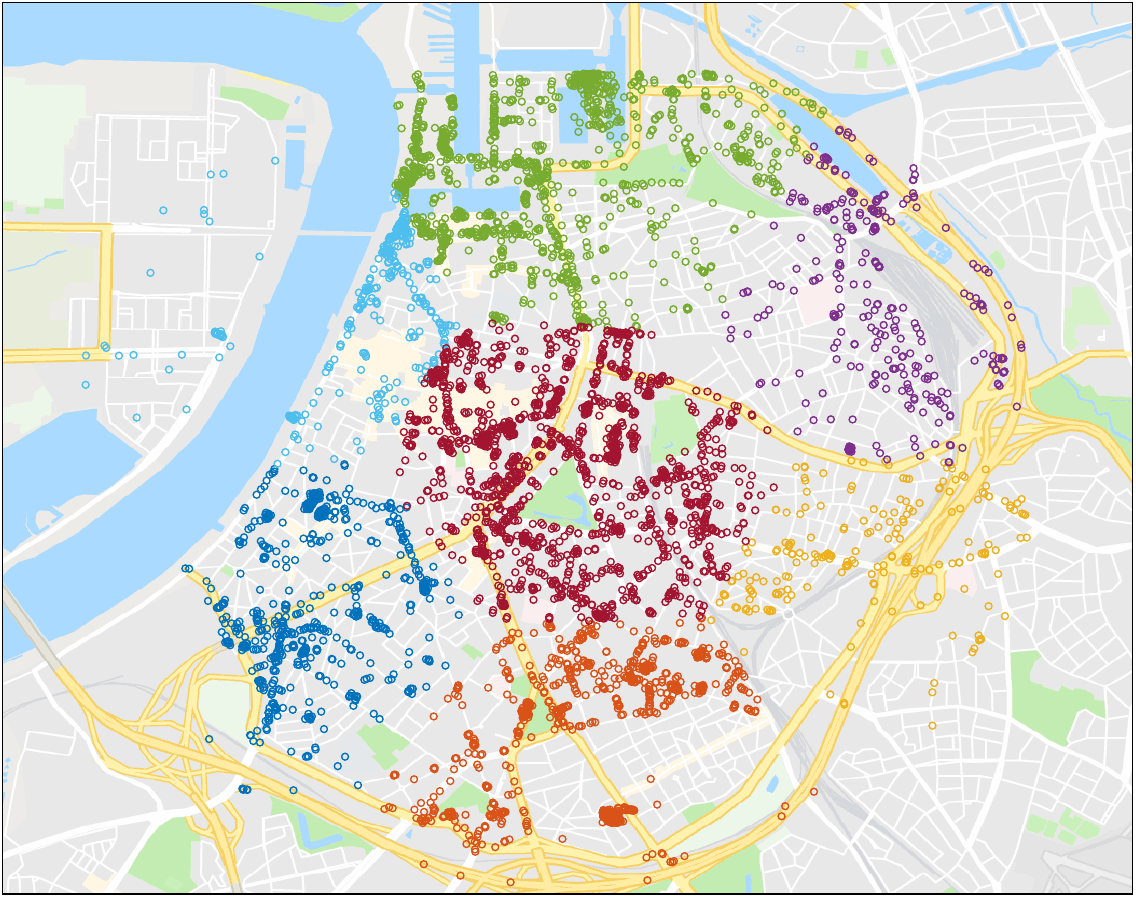}
		\caption{Seven classes}
		\label{7cmap}
	\end{subfigure}
	\begin{subfigure}[b]{0.33\textwidth}
		\includegraphics[width=\textwidth]{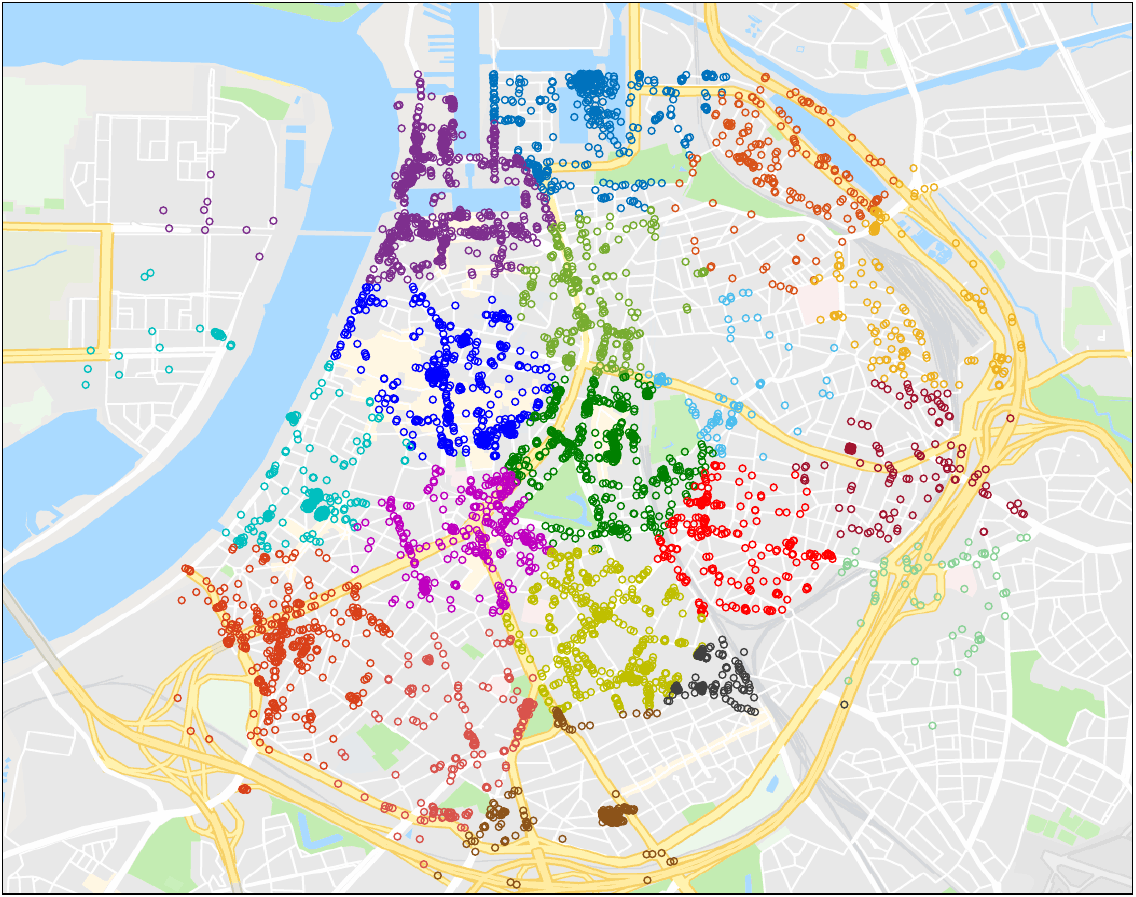}
		\caption{Eighteen classes}
		\label{18cmap}
	\end{subfigure}
	\caption{Maps showing the position of nodes divided into three, seven, and eighteen classes.}
	\label{cmap}
\end{figure*}

%===========================================================================
\section{Experimental Results}\label{results}
%===========================================================================
In this section, we present our experimental setup. Subsequently, results using IoT nodes that support communication over both Sigfox and \ac{TD}-LAN are detailed.

\subsection{Experiments setup}
%===========================================================================
The experiments have been conducted in two different scenarios, a university campus, and a city center.

\subsubsection{University campus}
In the second scenario, 16 nodes equipped with Sigfox modems are positioned in KU Leuven's Arenberg campus with $D \approx 2.5r$, as shown in Fig. \ref{map}. Each node sent 100 messages from which the \ac{RSSI} values are measured at the 3 \ac{BSs} that have received the messages. In Fig. \ref{map} the dark orange and dark blue nodes are \ac{$GSN$}s. Moreover,D2D communication is enabled over \ac{TD}-LAN network. Using D2D communication, we aim to further increase the localization accuracy within classes using the \ac{RSSI} measurement collected when nodes communicate with each other.
 
\subsubsection{City center}
Fig. \ref{cmap} presents the positions, in the city of Antwerp, from which Sigfox messages are collected \cite{antwerpDset}. In particular, over 14000 Sigfox messages were collected using 20 cars of the Belgian postal service \cite{antwerpDset} for two months and a half. Messages are sent from a region with a radius of 2.5\,km and are received by 58 Sigfox \ac{BSs}, in total. Figures \ref{3cmap}, \ref{7cmap}, and \ref{18cmap} show the virtual splitting of the city into 3, 7, and 18 classes with $D$ equals 1830\,m, 1600\,m, and 1300\,m, respectively. The dataset contains the GPS coordinates, base station ID, \ac{RSSI} of each Sigfox message. In order to resemble \ac{$GSN$} in this dataset, we assume that all measurements collected within a 100\,m radius from the center of each class represent one \ac{$GSN$}. This provides 65, 45, and 25 messages per class to be used as training data in the case of three, seven, and eighteen classes, respectively. Subsequently, the rest of the measurements are used as testing data to be classified. It is worth noting that, in our analysis, uniform test sets have been used by making sure that all training and test sets are of the same size for each class.

\subsection{Sigfox Communication - classification}
%===========================================================================
In this scenario, we investigate the localization accuracy using Sigfox network. In particular, we focus on a star IoT network scenario in which D2D is not available. In the following, we thoroughly investigate different design parameters that affect the classification accuracy. In this subsection, we mainly use the dataset collected in the city of Antwerp \cite{antwerpDset}, unless mentioned otherwise.

\begin{figure}[t]
	\centering
	\input{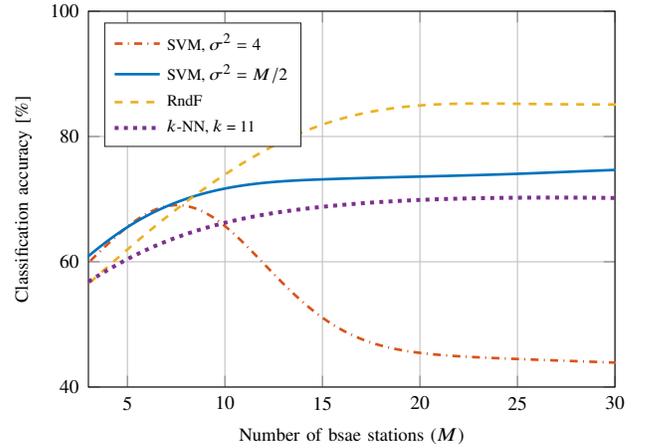}
	\caption{Classification accuracy versus the number of \ac{BSs} (features), 7 classes, $D$ = 1.6\,km, $r$ = 600\,m, and 40 training messages from each $GSN$ node.}
	\label{nrBS}
\end{figure}

\begin{figure}[t]
	\centering
	\includegraphics{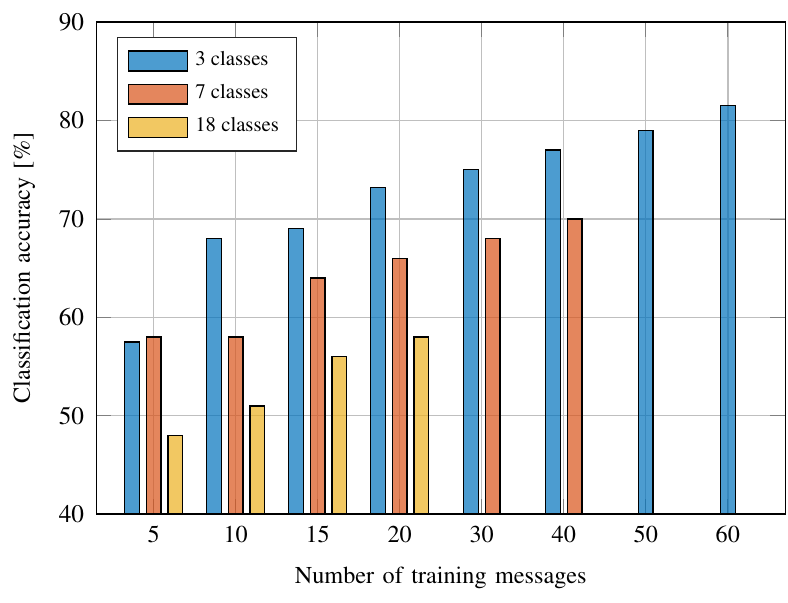}
	\caption{Classification accuracy versus the number of training messages from each $GSN$ node for three, seven, and eighteen classes with $D = \sqrt{3}r$.}
	\label{trainN}
\end{figure}

\begin{figure*}[t]
	\begin{subfigure}[b]{0.31\textwidth}
		\includegraphics[width=\textwidth]{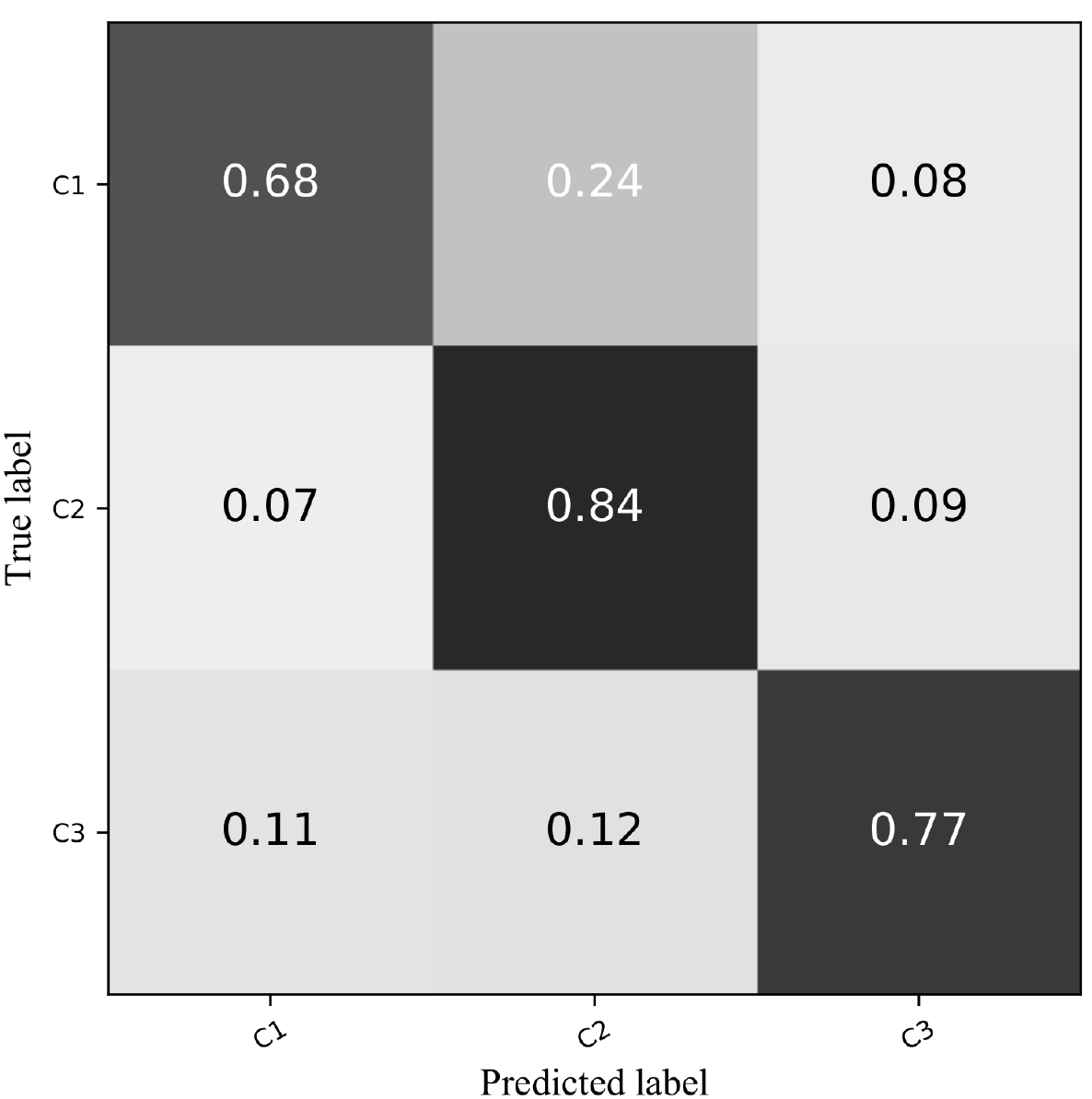}
		\caption{Three classes}
		\label{rforest500_3}
	\end{subfigure}
	\begin{subfigure}[b]{0.31\textwidth}
		\includegraphics[width=\textwidth]{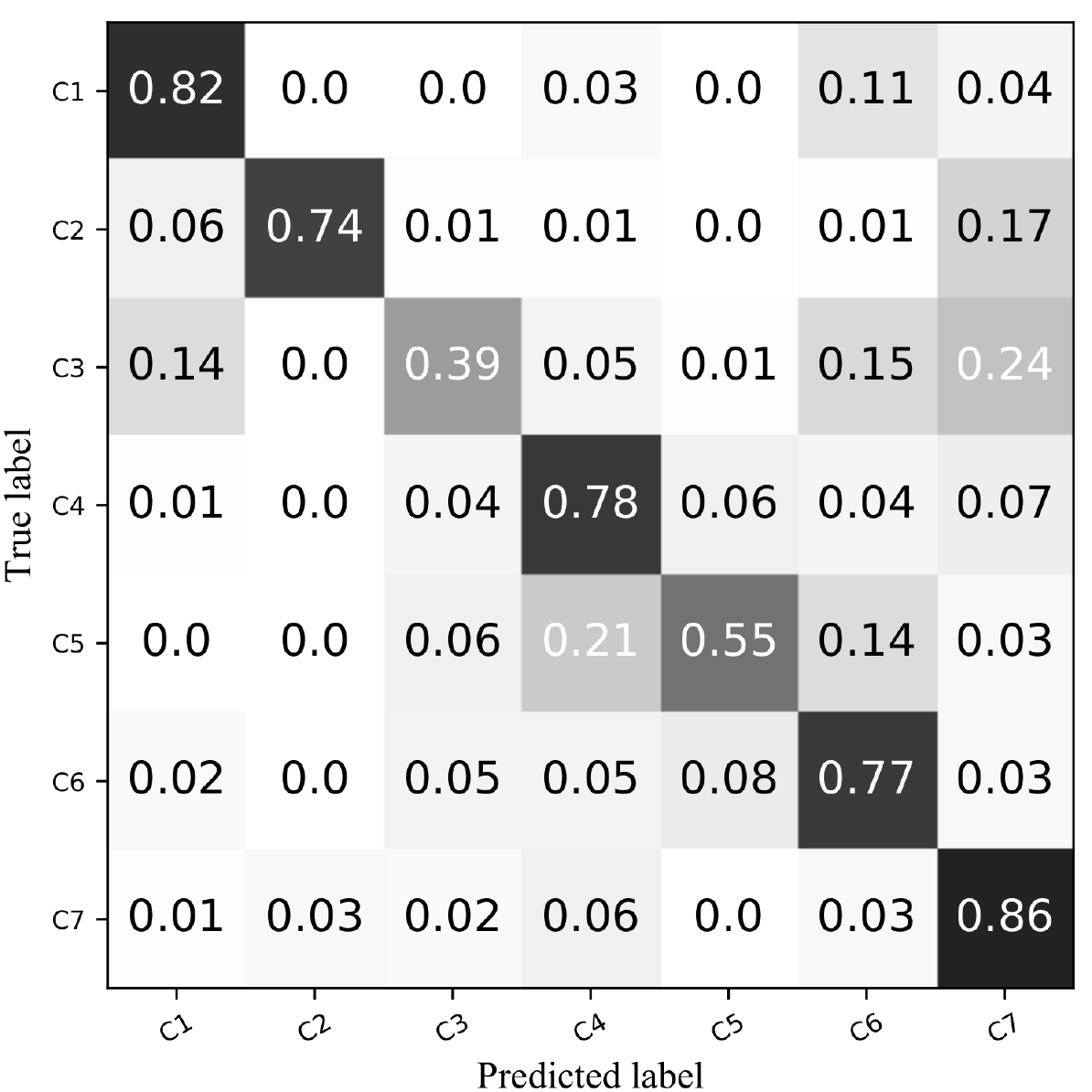}
		\caption{Seven classes}
		\label{rforest500_7}
	\end{subfigure}
	\begin{subfigure}[b]{0.36\textwidth}
		\includegraphics[width=\textwidth]{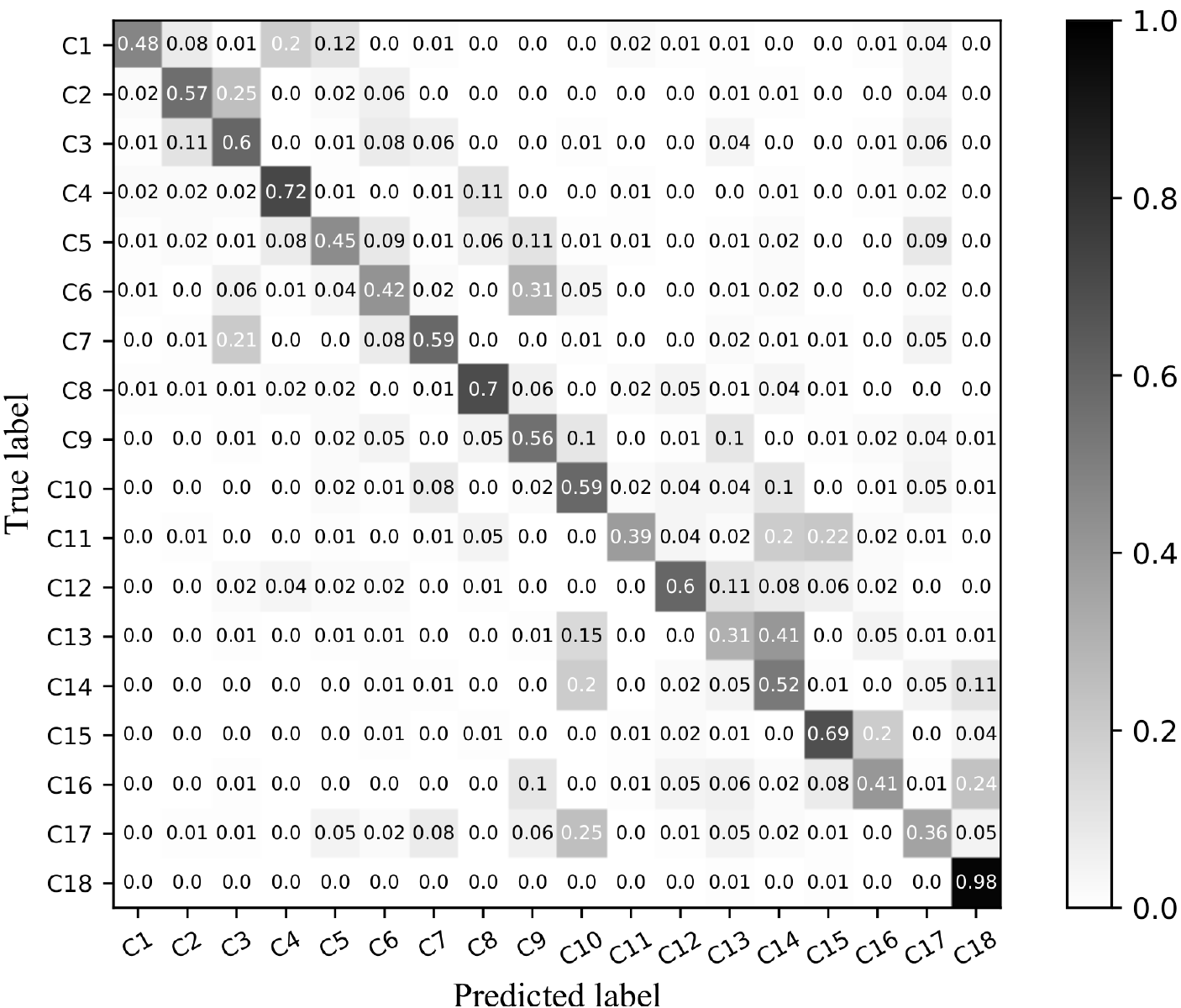}
		\caption{Eighteen classes}
		\label{rforest500_18}
	\end{subfigure}
	\caption{The confusion matrices for the case of three, seven, and eighteen adjacent classes with training data of 60, 40, and 20 messages from each $GSN$, respectively, and $D = \sqrt{3}r$.}
	\label{confMtrx}
\end{figure*}

\subsubsection{Number of features}
Fig. \ref{nrBS} presents the performance of \ac{RndF}, \ac{SVM}, and $k$-NN with a different number of features. As shown in the figure, using the proposed algorithm, \ac{RndF} outperforms $k$-NN presented in \cite{antwerpDset} with $k$ = 11. Moreover, it provides better performance than SVM when the number of features, namely, \ac{BSs}, is above 8. The reason behind this trend is twofold; first, it is well-known that SVM works better with separable datasets \cite{scikit, 2007overfitting}. However, in long-range UNB IoT network, increasing the number of features means adding \ac{BSs} that are more likely to be far away ($>5$\,km). Hence, having similar RSSI measurements makes the classes inseparable. \ac{RndF}, on the other hand, does not have such constraint, since similar features and samples are not used to extend trees. Secondly, SVM with a Gaussian kernel requires enough examples in order to find the optimal separating hyperplane in a higher dimensional space \cite{2007overfitting,Tran}. However, in Fig. \ref{nrBS}, the training messages are fixed to 40, from each \ac{$GSN$} node. The performance of \ac{SVM} and \ac{RndF} with a different number of training messages and a fixed number of features is presented in Fig. \ref{ave}.

Furthermore, Fig. \ref{nrBS} illustrates the effect of $\sigma^2$ on the performance of the \ac{SVM} classifier. As shown in the figure, selecting a fixed $\sigma^2 = 4$ results in lower classification accuracy with $M$, for $M > 8$. This is because SVM over-fits on the training data when the number of features is larger than 8. However, this over-fitting is avoided by using an adaptive $\sigma^2$ equals to $M/2$. In fact, the result presented in Fig. \ref{nrBS} extends the one reported in \cite{sallouha}, by giving deep insights on when and which \ac{ML} algorithm to use.

\subsubsection{Number of training messages}
In long-range IoT networks, the number of messages a node can send is limited. For instance, in Sigfox network each node can send 140 messages a day, with an average of one message every 10 minutes. Therefore, it is crucial to estimate the \ac{$SN$} class using as minimum training messages as possible. In Fig. \ref{trainN}, we present the performance of \ac{RndF} with three, seven, and eighteen classes using a different number of messages. As shown in the figure, having more training messages is beneficial for the classification accuracy. However, the trade-off here is the long time required to collect training messages. Nonetheless, it is worth noting that, collecting training messages is a cumulative process, therefore, the waiting time will be overcome in the long-term.

\subsubsection{Number of classes}
The number of classes, i.e., partitions, is a crucial parameter as it affects both the classification accuracy and localization accuracy. As shown in Figs. \ref{rforest500_3}, \ref{rforest500_7}, and \ref{rforest500_18}, using \ac{RndF} with 100 trees, for the case when the city is partitioned into three regions, the classification accuracy is much higher in comparison with the cases with seven and eighteen classes. However, the case with three classes provides less location information when compared with the case of eighteen classes. Accordingly, the trade-off is the classification accuracy versus the localization accuracy. In fact, this design parameter is mainly application dependent. Therefore, for applications such as packages tracking, it's sufficient enough to place \ac{$GSN$} on post offices checkpoints rather than placing extra \ac{$GSN$} all over the city. Consequently, better classification accuracy can be achieved.

\begin{figure}[t]
	\centering
	\input{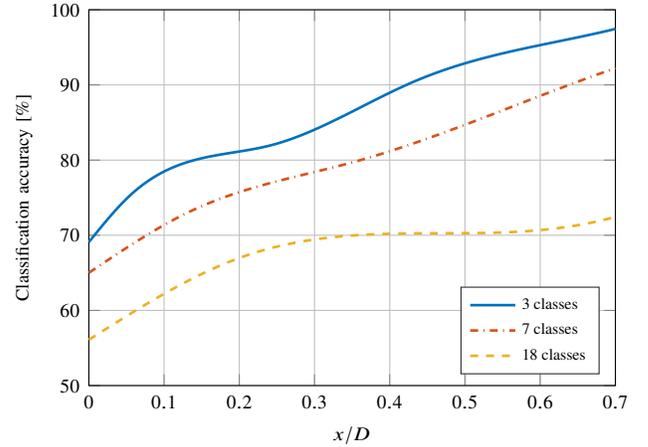}
	\caption{Classification accuracy using RndF with different $x/D$ and with 20 tanning messages from each $GSN$ node.}
	\label{Rradius}
\end{figure}

\subsubsection{Spacing between adjacent classes} %Classes radius
In order to quantify the impact of the spacing between adjacent classes, $x$, on the performance of the proposed algorithm, the classification accuracy as a function of $x/D$ is presented in Fig. \ref{Rradius}. Notice that $x \in [0,D)$, hence, dividing it by $D$ gives us the $x$-axis $\in [0,1)$. As shown in Fig. \ref{Rradius}, increasing the spacing between adjacent classes improves the classification accuracy, since the classes borders, which mostly cause classification errors, are eliminated. Hence, it is easier to distinguish messages. However, the cost here is the localization coverage region. In fact, since $D$ is fixed, increasing $x$ results in a region that is not fully included in the localization problem. Moreover, Fig. \ref{Rradius} illustrates the point at which the classes become connected, i.e., $D = \sqrt{3}r$. At this point, $x$ approaches zero and the classification accuracy is minimal as the number of \ac{$SN$}s at the classes' borders is maximal. Therefore, the minimum number of messages required to achieve a certain accuracy also increases. This is due to the fact that more training data is needed to learn the class of each node, particularly for \ac{$SN$}s at the border of the classes.

\subsubsection{Training time}
Algorithm \ref{alg:class} is set to retrain the model periodically with a retrain interval of one hour. Therefore, it is vital to investigate the training time needed. Considering the case of 7 classes, Table \ref{ttime} presents an estimate of the training time with a different number of training messages, from each $GSN$ node. In particular, Table \ref{ttime} shows that $k$-NN requires less time for training compared to both SVM and RndF. However, it is shown that the training time for both SVM and RndF is in the range of hundreds of milliseconds (ms). Such a training time is rather short in comparison with the message transmission rate in UNB IoT networks, which is one message per few minutes.

\begin{table}[!t]
	\small
	\caption{Training time of ML algorithms}
	\vspace{-0.5em}
	\label{ttime}
	\centering
	\begin{tabular}{|l||c|c|}
		\hline
		\textbf{Classifier} & \textbf{\# Train. messages} & \textbf{Train. time} [ms] \\ \hline
		\hline
		\multirow{2}{*}{RndF, 100 trees}                 & 40 &  200  \\ \cline{2-3} 
														 & 400 & 600  \\ \hline
		\multirow{2}{*}{RndF, 200 trees}                 & 40 &  400 \\ \cline{2-3} 
														 & 400 & 1200  \\ \hline
		\multirow{2}{*}{SVM}                             & 40 &  36 \\ \cline{2-3} 
														 & 400 & 460  \\ \hline
		\multirow{2}{*}{$k$-NN}                          & 40 &  10  \\ \cline{2-3} 
														 & 400 & 30 \\ \hline
	\end{tabular}
\end{table}

\begin{figure}[t]
	\centering
	% This file was created by matlab2tikz.
%
%The latest updates can be retrieved from
%  http://www.mathworks.com/matlabcentral/fileexchange/22022-matlab2tikz-matlab2tikz
%where you can also make suggestions and rate matlab2tikz.
%

\definecolor{mycolor1}{rgb}{0.00000,0.44700,0.74100}%
\definecolor{mycolor2}{rgb}{0.85000,0.32500,0.09800}%
\definecolor{mycolor3}{rgb}{0.92900,0.69400,0.12500}%
\begin{tikzpicture}

\begin{axis}[%
legend style={font=\fontsize{6}{5}\selectfont},
width=7cm,
height=5cm,
at={(1.154in,0.752in)},
scale only axis,
xmin=0,
xmax=70,
xmajorgrids,
xlabel={$\sigma{}^\text{2}$},
ymin=0.5,
ymax=0.9,
scaled y ticks={real:0.01},
ytick scale label code/.code={},
ymajorgrids,
ylabel={Classification accuracy [\%]},
axis background/.style={fill=white},
legend style={legend cell align=left,align=left,draw=white!15!black}
]
\addplot [color=mycolor1,line width=1.0pt]
  table[row sep=crcr]{%
0.1	0.52\\
0.125	0.535833333333333\\
0.166666666666667	0.56225\\
0.25	0.58375\\
0.333333333333333	0.6115\\
0.5	0.6405\\
0.666666666666667	0.65975\\
0.833333333333333	0.691\\
1	0.7135\\
1.25	0.72975\\
1.66666666666667	0.74375\\
2	0.75325\\
2.5	0.75965\\
3.33333333333333	0.75965\\
10	0.74515\\
15.1515151515152	0.70115\\
25	0.6529\\
33.3333333333333	0.60272\\
50	0.5312\\
71.4285714285714	0.5111\\
};
\addlegendentry{SVM, one-by-one};

\addplot [color=mycolor3,line width=1.0pt,dashed]
  table[row sep=crcr]{%
0.1	0.53\\
0.125	0.533333333333333\\
0.166666666666667	0.54675\\
0.25	0.56275\\
0.333333333333333	0.59275\\
0.5	0.63275\\
0.666666666666667	0.67275\\
0.833333333333333	0.714\\
1	0.75\\
1.25	0.778\\
1.66666666666667	0.796\\
2	0.808\\
2.5	0.8104\\
3.33333333333333	0.8044\\
10	0.7804\\
15.1515151515152	0.7253\\
25	0.6678\\
33.3333333333333	0.61162\\
50	0.5327\\
71.4285714285714	0.5111\\
};
\addlegendentry{SVM, averaged 5-by-5};

\addplot [color=mycolor2,line width=1.0pt,dashdotted]
  table[row sep=crcr]{%
0.1	0.53\\
0.125	0.5369\\
0.166666666666667	0.55011\\
0.25	0.56611\\
0.333333333333333	0.59847\\
0.5	0.63807\\
0.666666666666667	0.68065\\
0.833333333333333	0.7319\\
1	0.7819\\
1.25	0.8142\\
1.66666666666667	0.8358\\
2	0.848\\
2.5	0.8464\\
3.33333333333333	0.8244\\
10	0.78944\\
15.1515151515152	0.73034\\
25	0.66884\\
33.3333333333333	0.60666\\
50	0.5327\\
71.4285714285714	0.5111\\
};
\addlegendentry{SVM, averaged 10-by-10};

\end{axis}
\end{tikzpicture}%
	\caption{The classification accuracy as a function of $\sigma^2$ for the measurements conducted on the university campus.}
	\label{sigma}
\end{figure}
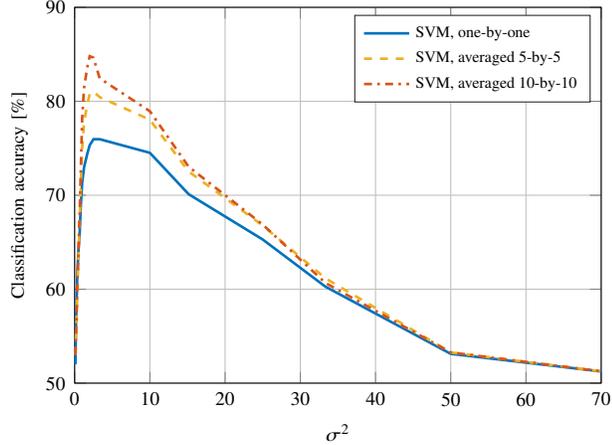

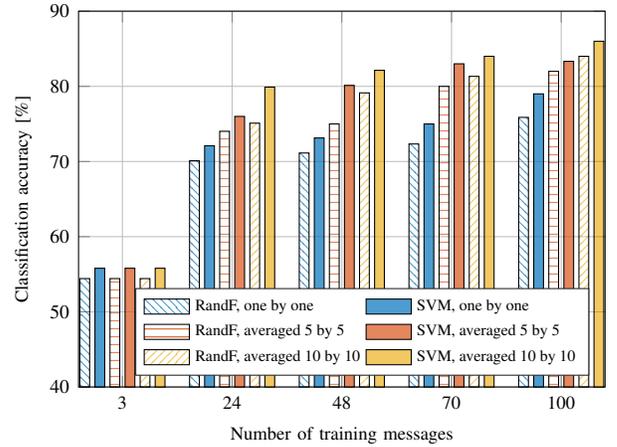
\begin{figure}[t]
	\centering
	% This file was created by matlab2tikz.
%
%The latest updates can be retrieved from
%  http://www.mathworks.com/matlabcentral/fileexchange/22022-matlab2tikz-matlab2tikz
%where you can also make suggestions and rate matlab2tikz.
%

\definecolor{mycolor1}{rgb}{0.00000,0.44700,0.74100}%
\definecolor{mycolor2}{rgb}{0.85000,0.32500,0.09800}%
\definecolor{mycolor3}{rgb}{0.92900,0.69400,0.12500}%

\begin{tikzpicture}

\begin{axis}[%
legend style={font=\fontsize{6}{5}\selectfont},
width=7cm,
height=5cm,
ybar,
bar width=0.13cm,
tick align=inside,
at={(0.758in,0.481in)},
scale only axis,
xmajorgrids,
symbolic x coords={3,24,48,70,100},
xtick=data,
xlabel={Number of training messages},
ymin=40,
ymax=90,
ymajorgrids,
ylabel={Classification accuracy [\%]},
axis background/.style={fill=white},
legend style={at={(0.97,0.03)},anchor=south east,legend cell align=left,align=left,draw=white!15!black,legend columns=2}
]
\addplot[fill=mycolor1,fill opacity=0.7,draw=black,area legend,pattern = north west lines,pattern color = mycolor1] plot table[row sep=crcr] {%
	x	y\\
	3	54.424\\
	24	70.102\\
	48	71.133\\
	70	72.333\\
	100	75.87\\
};
\addlegendentry{RandF, one by one};

\addplot[fill=mycolor1,fill opacity=0.7,draw=black,area legend] plot table[row sep=crcr] {%
	x	y\\
	3	55.8\\
	24	72.1\\
	48	73.133\\
	70	75\\
	100	79\\
};
\addlegendentry{SVM, one by one};

\addplot[fill=mycolor2,fill opacity=0.7,draw=black,area legend,pattern color = mycolor2, pattern=horizontal lines] plot table[row sep=crcr] {%
	x	y\\
	3	54.424\\
	24	74.02\\
	48	75\\
	70	80\\
	100	82\\
};
\addlegendentry{RandF, averaged 5 by 5};

\addplot[fill=mycolor2,fill opacity=0.7,draw=black,area legend] plot table[row sep=crcr] {%
	x	y\\
	3	55.8\\
	24	76\\
	48	80.133\\
	70	83\\
	100	83.33\\
};
\addlegendentry{SVM, averaged 5 by 5};

\addplot[fill=mycolor3,fill opacity=0.7,draw=black,area legend,pattern color = mycolor3, pattern = north east lines] plot table[row sep=crcr] {%
	x	y\\
	3	54.424\\
	24	75.102\\
	48	79.133\\
	70	81.333\\
	100	84\\
};
\addlegendentry{RandF, averaged 10 by 10};

\addplot[fill=mycolor3,fill opacity=0.7,draw=black,area legend] plot table[row sep=crcr] {%
	x	y\\
	3	55.8\\
	24	79.9\\
	48	82.133\\
	70	84\\
	100	86\\
};
\addlegendentry{SVM, averaged 10 by 10};

\end{axis}
\end{tikzpicture}%
	\caption{Classification accuracy of both SVM and RndF algorithms using the university campus dataset. The number of features is fixed and equals to $3$.}
	\label{ave}
\end{figure}

\subsubsection{RSSI averaging}
As shown in Fig. \ref{nrBS}, in case of a low number of features, \ac{SVM} outperforms RndF. Here, we focus on this particular case using the measurements conducted on the university campus. Recall that in university campus we have 3 \ac{BSs} and 100 messages. When \ac{SVM} is used, we first need to choose the value of $\sigma^2$ for the RBF kernel. In Fig. \ref{sigma}, we present the classification accuracy of 100 messages with $\sigma^2$. As shown in the figure, the best classification accuracy is achieved at $\sigma^2 \approx 4$. Accordingly, we choose a kernel function with $\sigma^2 = 4$ for our \ac{SVM}. Fig. \ref{ave} presents the effect of the size of the training data on the classification accuracy for both \ac{SVM} and RndF. As shown in the figure, SVM outperforms \ac{RndF} since enough training examples are available for the dataset with three features used. Maximum classification accuracy of 78\% is obtained when classifying messages one-by-one. However, by averaging the \ac{RSSI} values, for both $GSN$ and $SN$ nodes, we overcome the effect of small scale temporal fading. Hence, better accuracy is achieved. In particular, averaging the messages' \ac{RSSI} 10 by 10 brings the classification accuracy to 87\% using our \ac{SVM} classifier, as illustrated in Fig. \ref{ave}. The cost of this improvement is the delay required to receive ten messages before making an estimation.

A wrongly classified message implies that the estimated location of the \ac{$SN$} is in another class. Consequently, the corresponding distance error is the distance from the estimated class's center to the correct one. To further improve the accuracy, the impact of enabling D2D communication between nodes is presented in the following subsection.

\subsection{Device-to-Device enabled - Regression}
%===========================================================================

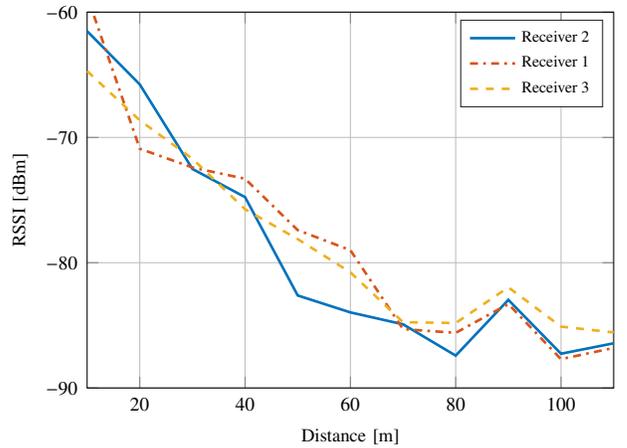
\begin{figure}[t]
	\centering
	% This file was created by matlab2tikz.
%
%The latest updates can be retrieved from
%  http://www.mathworks.com/matlabcentral/fileexchange/22022-matlab2tikz-matlab2tikz
%where you can also make suggestions and rate matlab2tikz.
%
\definecolor{mycolor1}{rgb}{0.00000,0.44700,0.74100}%
\definecolor{mycolor2}{rgb}{0.85000,0.32500,0.09800}%
\definecolor{mycolor3}{rgb}{0.92900,0.69400,0.12500}%
\begin{tikzpicture}

\begin{axis}[%
legend style={font=\fontsize{6}{5}\selectfont},
width=7cm,
height=5cm,
at={(1.154in,0.752in)},
scale only axis,
xmin=10,
xmax=110,
xmajorgrids,
xlabel={Distance [m]},
ymin=-90,
ymax=-60,
ymajorgrids,
ylabel={RSSI [dBm]},
axis background/.style={fill=white},
legend style={legend cell align=left,align=left,draw=white!15!black}
]
\addplot [color=mycolor1,solid,line width=1.0pt]
  table[row sep=crcr]{%
10	-61.5166666666667\\
20	-65.7666666666667\\
30	-72.5166666666667\\
40	-74.7666666666667\\
50	-82.6166666666667\\
60	-83.9666666666667\\
70	-84.9166666666667\\
80	-87.4166666666667\\
90	-82.9666666666667\\
100	-87.2666666666667\\
110	-86.4333333333333\\
};
\addlegendentry{Receiver 2};

\addplot [color=mycolor2,solid,line width=1.0pt,dashdotted]
  table[row sep=crcr]{%
10	-58.7\\
20	-70.9\\
30	-72.4\\
40	-73.3\\
50	-77.4\\
60	-79\\
70	-85.3\\
80	-85.6\\
90	-83.3\\
100	-87.7\\
110	-86.8\\
};
\addlegendentry{Receiver 1};

\addplot [color=mycolor3,solid,line width=1.0pt,dashed]
  table[row sep=crcr]{%
10	-64.7\\
20	-68.6333333333333\\
30	-71.7\\
40	-75.7333333333333\\
50	-78.1166666666667\\
60	-80.7666666666667\\
70	-84.75\\
80	-84.8166666666667\\
90	-81.9666666666667\\
100	-85.1\\
110	-85.5666666666667\\
};
\addlegendentry{Receiver 3};

\end{axis}
\end{tikzpicture}%
	\caption{Average RSSI values at three different receivers with a separation distance of 10\,m.}
	\label{3rec}
\end{figure}
\begin{figure}[t]
	\centering
	\input{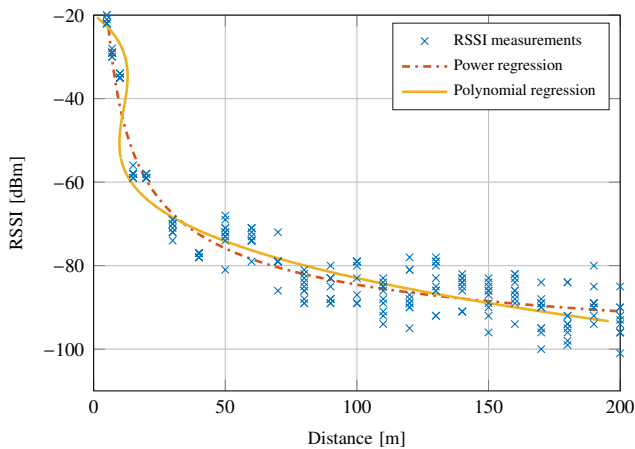}
	\caption{Power regression and polynomial regression of RSSI measurements collected in steps of 10\,m.}
	\label{regr}
\end{figure}
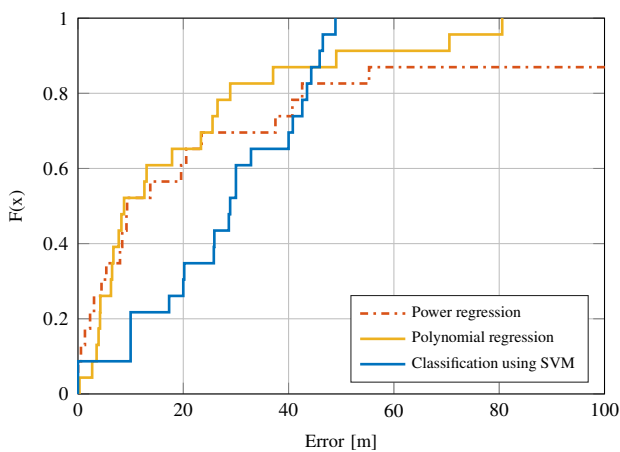
\begin{figure}[t]
	\centering
	% This file was created by matlab2tikz.
%
%The latest updates can be retrieved from
%  http://www.mathworks.com/matlabcentral/fileexchange/22022-matlab2tikz-matlab2tikz
%where you can also make suggestions and rate matlab2tikz.
%
\definecolor{mycolor1}{rgb}{0.00000,0.44700,0.74100}%
\definecolor{mycolor2}{rgb}{0.85000,0.32500,0.09800}%
\definecolor{mycolor3}{rgb}{0.92900,0.69400,0.12500}%
\begin{tikzpicture}

\begin{axis}[%
legend style={font=\fontsize{6}{5}\selectfont},
width=7cm,
height=5cm,
at={(1.387in,1.326in)},
scale only axis,
unbounded coords=jump,
xmin=0,
xmax=100,
xmajorgrids,
xlabel={Error [m]},
ymin=0,
ymax=1,
ymajorgrids,
ylabel={F(x)},
axis background/.style={fill=white},
title style={font=\bfseries},
legend style={at={(0.97,0.03)},anchor=south east,legend cell align=left,align=left,draw=white!15!black}
]
\addplot [color=mycolor2,line width=1.0pt,dashdotted]
  table[row sep=crcr]{%
-inf	0\\
0.00636472565428985	0\\
0.00636472565428985	0.0434782608695652\\
0.0233883126351838	0.0434782608695652\\
0.0233883126351838	0.0869565217391304\\
0.5674595849296	0.0869565217391304\\
0.5674595849296	0.130434782608696\\
1.32432190293137	0.130434782608696\\
1.32432190293137	0.173913043478261\\
2.29893555631424	0.173913043478261\\
2.29893555631424	0.217391304347826\\
3.04949488241802	0.217391304347826\\
3.04949488241802	0.260869565217391\\
4.47043295438757	0.260869565217391\\
4.47043295438757	0.304347826086957\\
5.38115622864427	0.304347826086957\\
5.38115622864427	0.347826086956522\\
7.98296700974063	0.347826086956522\\
7.98296700974063	0.391304347826087\\
8.40167586070025	0.391304347826087\\
8.40167586070025	0.434782608695652\\
9.21266198881331	0.434782608695652\\
9.21266198881331	0.478260869565217\\
9.33288666297229	0.478260869565217\\
9.33288666297229	0.521739130434783\\
13.73047293491	0.521739130434783\\
13.73047293491	0.565217391304348\\
19.5887604445114	0.565217391304348\\
19.5887604445114	0.608695652173913\\
20.5578763125345	0.608695652173913\\
20.5578763125345	0.652173913043478\\
23.4928935828165	0.652173913043478\\
23.4928935828165	0.695652173913043\\
37.500656557522	0.695652173913043\\
37.500656557522	0.739130434782609\\
40.7209778496085	0.739130434782609\\
40.7209778496085	0.782608695652174\\
42.5776844272856	0.782608695652174\\
42.5776844272856	0.826086956521739\\
55.2888891060046	0.826086956521739\\
55.2888891060046	0.869565217391304\\
132.079076577927	0.869565217391304\\
132.079076577927	0.91304347826087\\
164.070978813975	0.91304347826087\\
164.070978813975	0.956521739130435\\
170.625004750871	0.956521739130435\\
170.625004750871	1\\
inf	1\\
};
\addlegendentry{Power regression};

\addplot [color=mycolor3,line width=1.0pt]
  table[row sep=crcr]{%
-inf	0\\
0.290404907000124	0\\
0.290404907000124	0.0434782608695652\\
2.69457957899996	0.0434782608695652\\
2.69457957899996	0.0869565217391304\\
3.54067703200002	0.0869565217391304\\
3.54067703200002	0.130434782608696\\
3.91329900799983	0.130434782608696\\
3.91329900799983	0.173913043478261\\
4.18574699999996	0.173913043478261\\
4.18574699999996	0.217391304347826\\
4.24918088299998	0.217391304347826\\
4.24918088299998	0.260869565217391\\
6.26658258100017	0.260869565217391\\
6.26658258100017	0.304347826086957\\
6.47980844800023	0.304347826086957\\
6.47980844800023	0.347826086956522\\
6.72793369599978	0.347826086956522\\
6.72793369599978	0.391304347826087\\
7.74448589900001	0.391304347826087\\
7.74448589900001	0.434782608695652\\
8.24868411700001	0.434782608695652\\
8.24868411700001	0.478260869565217\\
8.75395513599969	0.478260869565217\\
8.75395513599969	0.521739130434783\\
12.6146161039999	0.521739130434783\\
12.6146161039999	0.565217391304348\\
13.033184099	0.565217391304348\\
13.033184099	0.608695652173913\\
17.868712	0.608695652173913\\
17.868712	0.652173913043478\\
23.3582080000001	0.652173913043478\\
23.3582080000001	0.695652173913043\\
25.5570937519998	0.695652173913043\\
25.5570937519998	0.739130434782609\\
26.5043609089997	0.739130434782609\\
26.5043609089997	0.782608695652174\\
28.9125575679999	0.782608695652174\\
28.9125575679999	0.826086956521739\\
37.0544347870001	0.826086956521739\\
37.0544347870001	0.869565217391304\\
49.0503781360001	0.869565217391304\\
49.0503781360001	0.91304347826087\\
70.538559621	0.91304347826087\\
70.538559621	0.956521739130435\\
80.5708950009998	0.956521739130435\\
80.5708950009998	1\\
inf	1\\
};
\addlegendentry{Polynomial regression};
\addplot [color=mycolor1,solid,line width=1.0pt]
table[row sep=crcr]{%
	-inf	0\\
	0	0\\
	0	0.0869565217391304\\
	10	0.0869565217391304\\
	10	0.217391304347826\\
	17.31	0.217391304347826\\
	17.31	0.260869565217391\\
	20	0.260869565217391\\
	20	0.304347826086957\\
	20.18	0.304347826086957\\
	20.18	0.347826086956522\\
	25.8	0.347826086956522\\
	25.8	0.391304347826087\\
	25.9	0.391304347826087\\
	25.9	0.434782608695652\\
	28.65	0.434782608695652\\
	28.65	0.478260869565217\\
	28.9	0.478260869565217\\
	28.9	0.521739130434783\\
	30	0.521739130434783\\
	30	0.608695652173913\\
	32.86	0.608695652173913\\
	32.86	0.652173913043478\\
	40	0.652173913043478\\
	40	0.695652173913043\\
	40.8	0.695652173913043\\
	40.8	0.739130434782609\\
	42.6	0.739130434782609\\
	42.6	0.782608695652174\\
	43.5	0.782608695652174\\
	43.5	0.826086956521739\\
	44.3	0.826086956521739\\
	44.3	0.869565217391304\\
	45.9	0.869565217391304\\
	45.9	0.91304347826087\\
	46.5	0.91304347826087\\
	46.5	0.956521739130435\\
	48.9	0.956521739130435\\
	48.9	1\\
	inf	1\\
};
\addlegendentry{Classification using SVM};

\end{axis}
\end{tikzpicture}%
	\caption{Empirical CDF: performance of regression vs. fingerprinting for the measurements conducted on the university campus.}
	\label{pwrVspol}
\end{figure}

When the density of \ac{$GSN$} nodes is high enough, and in case \ac{TD}-LAN is enabled, \ac{RSSI} can be measured in the near zone where the \ac{RSSI}-distance curve has sufficient resolution. To estimate the distance using \ac{RSSI}, we first collect discrete measurements and subsequently, use a regression process to fit \ac{RSSI}-distance curve. To this end, we used three nodes in receiving mode, placed on a line with 10\,m inter-distance. Then, we placed transmitters on a line, perpendicular to the line made by the receivers, at distances from 10\,m to 200\,m with a spacing of 10\,m. It is worth mentioning that, this setup is not optimal. An optimal setup requires receiving nodes to be placed on the corners of the considered area to have positions diversity. Hence, the obtained accuracy can be considered as a lower bound. The measurements of this setup are illustrated in Fig. \ref{3rec}. The interpolation of \ac{RSSI} measurements using power series and polynomial regression is shown in Fig. \ref{regr}. In particular, we used a two-term power series regression and a third-degree polynomial regression. This results in the curves shown in Fig. \ref{regr}, which are used for estimating the distances in the test phase.

Finally, Fig. \ref{pwrVspol} presents the cumulative distribution function (CDF) of the localization error for both power series and polynomial regression. It can be concluded that similar performance is obtained from both methods. Furthermore, for the same setup, the performance of fingerprinting localization is presented. While distance estimation outperforms fingerprinting localization for errors lower than 40\,m, fingerprinting localization error is always bounded by the distances between \ac{$GSN$}s. Distance estimation at distances further than 200\,m from the receiver results in high errors ($>$60\,m). Therefore, it is recommended to limit the use of RSSI ranging-based localization to classes with a maximum radius of 200\,m.

%===========================================================================
\section{Conclusion}\label{conc}
%===========================================================================
This paper presents localization methods for \ac{UNB} IoT networks by exploiting the available \ac{$GSN$}s to estimate the location of other \ac{$SN$}s. Firstly, \ac{RSSI} measurements from \ac{$GSN$}s are used as a source of training data to classify other \ac{$SN$} into location-based classes. By using only communication over Sigfox, this solution offers the localization service for free, since \ac{$GSN$}s have to send their own messages anyway. Secondly, it has been shown that using D2D communication, regression-based distance estimation improves the localization accuracy of the classified \ac{$SN$}s. However, this requires extra \ac{$GSN$} nodes. Using measurement based dataset, we introduced the design parameters which affect the performance of the proposed approach, such as the number of classes, the spacing between adjacent classes and the number of features used. Furthermore, the effect of each of these parameters, along with the trade-offs introduced with them, has been thoroughly investigated.

\section*{Acknowledgment}
The authors would like to thank the Sigfox operator in Belgium, Engie--M2M, for their cooperation. 

\balance
\bibliographystyle{ieeetr} 
\bibliography{iot-IEEEbib}

\end{document}